\pdfoutput=1
\documentclass[
    twocolumn,
	prd,
	amssymb,
	preprintnumbers,
	secnumarabic,
	nofootinbib]{revtex4-1}

\usepackage{graphicx}
\usepackage{enumitem}
\usepackage{latexsym}
\usepackage{amsfonts}
\usepackage{amssymb}
\usepackage{color}
\usepackage{xcolor}
\usepackage{amsmath}
\usepackage{bm}
\usepackage{slashed}
\usepackage{dcolumn}
\usepackage{verbatim}

\usepackage{multirow}
\usepackage{xspace}
\usepackage[normalem]{ulem}
\usepackage[
pdfauthor={Manish Tamta}]{hyperref}

\definecolor{midgreen}{rgb}{0.0,0.675,0.0}
\definecolor{aquamarine}{rgb}{0.2,0.7,0.6}
\definecolor{cerulean}{RGB}{0,166,214} 
\definecolor{subtlered}{rgb}{0.8,0.3,0.3}

\newcommand{\Ml}{M_{\rm L}}

\hypersetup{
 pdfauthor={Manish Tamta, Miguel Zumalacárregui},
  colorlinks=true,
  citecolor=cerulean,
  urlcolor=aquamarine,
  linkcolor=cerulean
}

\allowdisplaybreaks

\usepackage{scalerel,tikz}
\usetikzlibrary{svg.path}
\definecolor{orcidlogocol}{HTML}{A6CE39}
\tikzset{orcidlogo/.pic={
 \fill[orcidlogocol] svg{M256,128c0,70.7-57.3,128-128,128C57.3,256,0,198.7,0,128C0,57.3,57.3,0,128,0C198.7,0,256,57.3,256,128z};
 \fill[white] svg{M86.3,186.2H70.9V79.1h15.4v48.4V186.2z}
 svg{M108.9,79.1h41.6c39.6,0,57,28.3,57,53.6c0,27.5-21.5,53.6-56.8,53.6h-41.8V79.1z M124.3,172.4h24.5c34.9,0,42.9-26.5,42.9-39.7c0-21.5-13.7-39.7-43.7-39.7h-23.7V172.4z}
 svg{M88.7,56.8c0,5.5-4.5,10.1-10.1,10.1c-5.6,0-10.1-4.6-10.1-10.1c0-5.6,4.5-10.1,10.1-10.1C84.2,46.7,88.7,51.3,88.7,56.8z};
}}
\newcommand\orcidicon[1]{\href{https://orcid.org/#1}{\mbox{\scalerel*{
\begin{tikzpicture}[yscale=-1,transform shape]
\pic{orcidlogo};
\end{tikzpicture}
}{|}}}}

\begin{document}

\title{Fuzzy Dark Matter as a diffractive barrier: \\ Wave-optics Signatures of Ultralight Fields on Lensed Gravitational Waves}

\author{Manish Tamta$^{\orcidicon{0009-0006-7088-8705}}$} \email{manishtamta@iisc.ac.in}
\affiliation{Centre for High Energy Physics, Indian Institute of Science, C. V. Raman Avenue, Bengaluru 560012, India}

\author{Han Gil Choi$^{\orcidicon{0000-0002-1390-9024}}$} \email{hangil.choi@aei.mpg.de}
\affiliation{Max Planck Institute for Gravitational Physics (Albert Einstein Institute) \\ Am M\"uhlenberg 1, D-14476 Potsdam-Golm, Germany}

\author{Diego Blas$^{\orcidicon{0000-0003-2646-0112}}$}
\email{dblas@ifae.es}
\affiliation{Institut de F\'isica d'Altes Energies (IFAE), The Barcelona Institute of Science and Technology, 08193 Bellaterra (Barcelona), Spain}
\affiliation{Instituci\'o Catalana de Recerca i Estudis Avan\c{c}ats (ICREA), Passeig Llu\'is Companys 23, Barcelona, 08010, Spain}

\author{Miguel Zumalac\'arregui$^{\orcidicon{0000-0002-9943-6490}}$} \email{miguel.zumalacarregui@csic.es}
\affiliation{Instituto de F\'isica Te\'orica UAM/CSIC
C/ Nicol\'as Cabrera 13-15
Universidad Aut\'onoma de Madrid
Cantoblanco, Madrid 28049, Spain}
\affiliation{Max Planck Institute for Gravitational Physics (Albert Einstein Institute) \\ Am M\"uhlenberg 1, D-14476 Potsdam-Golm, Germany}

\date{\today}

\begin{abstract}
Gravitational waves (GWs) propagate through the universe from sources at high redshift, experiencing the gravitational fields of the intervening matter distribution. Halos composed of ultralight dark matter (ULDM) contain strong density fluctuations, reflecting the interference pattern in the field distribution of the ultralight bosons. In the wave-optics regime, ULDM fluctuations act as a diffractive grating, imprinting frequency-dependent modulations of GWs, in addition to magnification and other gravitational lensing effects. 
Since lensing depends on the projected surface density, we derive its approximate probability distribution and show how it departs from a Gaussian distribution depending on the halo profile and coherence length. We then quantify how the GW diffraction signatures produced by the granules depend on macroimage magnification and parity.
The ULDM interference network imprints stochastic diffraction fluctuations at the $\mathcal{O}(1)\%$ level on GWs moderately lensed by the host when the Fresnel scale is comparable to the granule coherence length $r_F(f)\sim\ell_c$. These fluctuations can be enhanced to $\gtrsim10\%$ when the host macroimage forms near the critical curve or for Type~II macroimages. For massive black hole binaries observed by LISA, the diffraction signature can be detectable around the ULDM particle mass $m_\phi\sim10^{-21}$~eV, providing either a smoking-gun signature of ULDM or new constraints on $m_\phi$.
\end{abstract}

\preprint{IFT-UAM/CSIC-26-153}

\maketitle

\tableofcontents

%%%%%%%%%%%%%%%%%%%%%%%%%
\section{Introduction}
\label{sec:intro}
%%%%%%%%%%%%%%%%%%%%%%%%%

Ultralight dark matter (ULDM), ``wave'' dark matter, is a dark matter model where the latter is represented by a collection of bosons with mass $m_\phi \lesssim \mathcal{O}(\mathrm{eV})$, whose coherent oscillations behave as non-relativistic matter~\cite{Turner:1983he, HuBarkanaGruzinov2000} (see Refs.~\cite{Marsh:2015xka,HuiOstrikerTremaineWitten2017,Niemeyer:2019aqm,Hui:2021tkt,Ferreira:2020fam} for reviews). Such a light, weakly coupled boson is a well-motivated dark-matter candidate, arising generically as a pseudo Nambu–Goldstone mode of a spontaneously broken approximate global symmetry, as for the axion-like fields of string compactifications~\cite{Svrcek:2006yi,Arvanitaki:2009fg}.

The most extreme masses, or around $m_\phi \sim \mathcal{O}(10^{-21}\mathrm{eV})$, are also known as 
``fuzzy'' dark matter, and was proposed as an alternative to cold dark matter (CDM) that suppresses structure on small scales while remaining observationally indistinguishable from CDM on larger scales~\cite{Baldeschi:1983mq,Khlopov:1985fch,Press:1989id,Sin:1992bg,Ji:1994xh,Lee:1995af,Matos:1998vk,Sahni:1999qe,HuBarkanaGruzinov2000,Goodman:2000tg,Peebles:2000yy}, as later confirmed by Schr\"odinger--Poisson simulations of cosmic structure~\cite{Widrow:1993qq,Woo:2008nn,Schive:2014dra, Schive:2014hza}. The reason behind this phenomenon is that, for a galactic halo with velocity dispersion $\sigma_v \sim 100~\mathrm{km/s}$, its typical wavelength is on the order of tens of parsecs, which complicates the gravitational contraction at smaller scales. The ULDM mass is constrained by the Lyman-$\alpha$ forest~\cite{Irsic:2017yje,Armengaud:2017nkf,Rogers:2020ltq} and by the sizes and stellar kinematics of ultrafaint dwarf galaxies~\cite{Dalal:2022rmp}.

The same wave nature makes a virialized ULDM halo granular. Interference among many wave modes that compose the field generates order-unity density fluctuations--``granules'' superposed on a smooth halo profile, with a coherence length set by the local de Broglie wavelength~\cite{HuiOstrikerTremaineWitten2017}, as seen in simulations~\cite{Schive:2014dra, Mocz:2017wlg,Veltmaat:2018dfz}; away from the central soliton, the field is well described as a random superposition of these waves~\cite{Bar-Or:2018pxz,Hui:2020hbq,Kawai:2021znq}, while at the center they condense into a coherent structure \cite{Schive:2014hza}. This granularity is a generic prediction of ULDM, and it imprints a statistical signature on any signal that propagates through the halo. The same fluctuations can be probed through their direct gravitational coupling to interferometers~\cite{Kim2023Interferometers}, pulsar timing residuals~\cite{KimMitridate2023PTA}, the astrometric deflection of background sources~\cite{Kim2024Astrometry}, and precise study of gravitationally bound systems~\cite{Blas2024Heterodyne,Blas:2016ddr}. Gravitational lensing, which responds to mass alone, offers a complementary and particularly clean probe.

Electromagnetic lensing already constrains ULDM granularity. Strong lensing has long been used to detect dark substructure, through flux-ratio anomalies of lensed quasars and the gravitational imaging of extended arcs~\cite{Dalal:2001fq,Vegetti:2009cz,Vegetti:2012mc,Hsueh:2019ynk,Gilman:2019nap,Sengul:2021lxe,Sengul:2022edu,Keeley:2024brx,Ephremidze:2025mqg}. For ULDM, the lensing properties of solitonic halos~\cite{Herrera-Martin:2017cux,Shevchuk:2023ccb} and the flux-ratio, astrometric and time-delay anomalies induced by granules~\cite{Chan2020FDMCriticalCurves,Laroche:2022pjm,Liu:2024uvr,Zhou2026WaveSimFDMLensing} have been studied. Very long baseline interferometry of an extended lensed radio arc excludes $m_\phi\lesssim 4.4\times10^{-21}$~eV~\cite{Powell2023VLBI_FDM}; anomalies in the image structure of an Einstein ring have also been used to try to distinguish wave-like dark matter from standard cold dark matter (CDM)~\cite{Amruth2023EinsteinRings}. Microlensing of stars extremely magnified by galaxy clusters is sensitive to the same granularity through a different channel, corrugation of the macrocaustic rather than image morphology~\cite{Venumadhav2017Microcaustics,Dai:2020rio,Broadhurst:2024ggk,Muller2026}.
Overall,  strong lensing effects offer a mature probe of dark matter substructure on the scales affected by ULDM~\cite{Vegetti2023Review}.

The gravitational-wave (GW) analogue is qualitatively different. For the halo masses and frequencies of interest here, lensing falls in the wave-optics regime~\cite{Ohanian:1974ys,Bontz:1981rvr,DeguchiWatson1986,Nakamura:1997sw,NakamuraDeguchi1999,TakahashiNakamura2003}, and the physics is that of a wave crossing a diffractive barrier. A screen whose structure varies on a scale $\ell$ affects a passing wave in three distinct ways, depending on how $\ell$ compares to the Fresnel scale $r_F\sim\sqrt{\lambda D_{\rm eff}}$, the size of the region that contributes coherently to the signal at wavelength $\lambda$~\cite{1992grle.book.....S}. Here $D_{\rm eff}=D_LD_{LS}/D_S$ with $D_L$, $D_S$ and $D_{LS}$ the angular-diameter distances to the lens, to the source and between them.
If $\ell\gg r_F$, rays are well defined and geometric optics applies; if $\ell\ll r_F$, the wave averages over many structures within a Fresnel zone and is essentially unaffected; if $\ell\sim r_F$, the wave diffracts, and the lensed signal can no longer be interpreted as a superposition of copies propagating along well-defined paths. The observable is then a complex, frequency-dependent amplification factor~\cite{TakahashiNakamura2003,Tanaka:2023mvy,Suyama:2025gbh}, which encodes the interference of paths that geometric optics would either separate into distinct images or fail to resolve at all. GW detectors record the phase-coherent waveform and hence measure this amplification factor directly; its frequency dependence, unlike a constant rescaling of the amplitude or shift of the arrival time, is not fully degenerate with the source parameters~\cite{Dai:2018enj,Tambalo2023HaloProbe,Savastano2023WOF}.

Lenses of astrophysical mass diffract GWs, whose wavelengths are long, and leave their imprint on the observed signal itself, without the need to resolve or separate lensed images. The amplification factor of a single lens encodes its mass and density profile, which has motivated wave-optics lensing as a probe of compact objects, dark-matter halos and subhalos, and extended dark objects~\cite{TakahashiNakamura2003,Dai:2018enj,Choi2021SmallScaleShear,Tambalo2023HaloProbe,Fairbairn:2022xln,Choi2024CoexistencePBH,Blinov:2026bxj,Choi:2026sjq}. Diffraction is strongest when the wave period is comparable to the light-crossing time of the lens mass enclosed within the impact parameter, i.e. when the lens enclosed mass satisfies $\Ml\sim10^4\,M_\odot\,(f/{\rm Hz})^{-1}$~\cite{TakahashiNakamura2003,Dai:2018enj}, so each frequency band probes a different mass range. At $f\sim10$--$10^3$~Hz, current ground-based detectors, LIGO, Virgo and KAGRA~\cite{LIGOScientific:2014pky,VIRGO:2014yos,KAGRA:2020tym}, and the next-generation Einstein Telescope~\cite{ET:2025xjr} and Cosmic Explorer~\cite{Evans:2021gyd} are sensitive to $\Ml\sim10$--$10^3\,M_\odot$: compact objects and the central regions of minihalos~\cite{Jung:2017flg,Dai:2018enj,Guo:2022dre,cheung2024probingminihalolensesdiffracted}. Proposed decihertz detectors, including lunar antennas~\cite{Ajith:2024mie,Jani:2020gnz,Jani:2025uaz}, atom interferometers~\cite{Badurina:2019hst,AEDGE:2019nxb} and space-borne interferometers such as DECIGO~\cite{Kawamura:2020pcg} and the Decihertz Observatory~\cite{Sedda:2021yhn} (see Ref.~\cite{Perego:2025bif} for a comparison of post-LISA concepts), would extend the reach to $\Ml\sim10^3$--$10^5\,M_\odot$~\cite{Guo:2022dre}. At millihertz frequencies, LISA~\cite{Colpi2024LISADSR} and Taiji~\cite{Hu:2017mde}, together with TianQin~\cite{TianQin:2015yph}, whose shorter arms shift its sensitivity to somewhat higher frequencies, will observe massive black hole binaries diffracted by $\Ml\sim10^5$--$10^8\,M_\odot$, the central regions of dark-matter halos and subhalos~\cite{Tambalo2023HaloProbe,Caliskan:2023zqm,Savastano2023WOF,Brando2024WeaklyLensedGW}, and proposed microhertz missions such as $\mu$Ares~\cite{Sesana:2019vho} and GUEST~\cite{Blas:2026xol} would reach $\Ml\gtrsim10^9\,M_\odot$, the scale of galaxies.

Forecasts predict that a small fraction of the binaries observed by ground-based detectors are \emph{strongly lensed} by galaxies and clusters~\cite{Ng:2017yiu,Li:2018prc,Oguri:2018muv,Xu:2021bfn,Wierda:2021upe,Mukherjee:2021qam,Smith:2022vbp}, rising to many events per year for next-generation observatories~\cite{Biesiada:2014kwa,Ding:2015uha,ET:2025xjr}, so the first confident detections are likely to come soon (see Refs.~\cite{Liao:2022gde,Grespan:2023cpa,Smith:2025axx} for reviews). Diffraction by stars in the lens galaxy distorts strongly lensed signals and can itself serve as evidence of strong lensing~\cite{Diego:2019lcd,Mishra:2021xzz,Shan:2023ngi,Shan:2023qvd,Zumalacarregui:2026uqs}. Searches for lensing signatures, which can exploit the Morse phase that distinguishes the image types~\cite{Dai:2017huk,Haris:2018vmn,Ezquiaga:2020gdt,Wang:2021kzt,Janquart:2021nus}, have been carried out both by independent groups~\cite{Dai:2020tpj,Chakraborty:2025maj,Chakraborty:2026yci} and by the LIGO--Virgo--KAGRA collaboration, whose analyses now routinely include wave-optics effects. No confident lensed event has been found so far~\cite{LIGOScientific:2021izm,LIGOScientific:2023bwz,LIGOScientific:2025cwb}, and GW231123 has been examined for diffraction signatures~\cite{LIGOScientific:2025cwb,Goyal:2025eqo,Shan:2025dcd,Chan:2025kyu,Chakraborty:2025pxt,Cheung:2026pky}. In the LISA band, a fraction of massive black hole binaries are also expected to be strongly lensed~\cite{Sereno:2010dr,Sereno:2011ty,Caliskan2023LISAObservability,Gutierrez:2025ymd}, and their high signal-to-noise ratios and long durations make them ideal targets for wave-optics studies~\cite{LISACosmologyWorkingGroup:2022jok}.

Beyond individual lenses, a GW propagating through the Universe is diffracted by the many unresolved structures along its path. Each perturbs the wave only slightly, but together they produce a random, frequency-dependent modulation of the amplitude and phase of every signal~\cite{Macquart:2004sh,Takahashi2006AmpPhase,Oguri:2020ldf,Inamori:2021tlx,Mizuno:2023pru,Urrutia:2024pos,Zumalacarregui2024LSD,Yarimoto:2024uew,Tanaka:2025nfw,Choi2026LISAStochasticDiffraction}. The statistics of this \emph{stochastic diffraction}~\cite{Zumalacarregui2024LSD,Choi2026LISAStochasticDiffraction} depend not on the properties of any single lens, but on the abundance of compact objects and on the small-scale matter power spectrum~\cite{Oguri:2020ldf,Kim:2025njb}, precisely the scales on which dark-matter models may differ.

Because the contributions of the many lenses compound, stochastic diffraction increases the sensitivity to the existence of lenses for GWs (the diffraction variances of many sightlines add), and can probe far lighter lenses than single-lens diffraction~\cite{Zumalacarregui2024LSD}. Ground-based and decihertz detectors could probe low-mass halos, axion minihalos and primordial black holes down to $\sim10^{-3}\,M_\odot$~\cite{Oguri:2020ldf,Kim:2025njb}. In the LISA band, standard cold dark matter (CDM) halos of $10$--$10^4\,M_\odot$, orders of magnitude below the single-lens range, leave a sub-percent stochastic fluctuation on every signal, which will allow LISA to constrain or even discover them~\cite{Choi2026LISAStochasticDiffraction}; eccentric binaries extend the reach to halos $\lesssim10\,M_\odot$~\cite{Urrutia:2024pos}. The effect is enhanced in strongly lensed events, where the external potential of the host magnifies the response to small-scale perturbations~\cite{OguriTakahashi2022AmpPhase,Nakazono:2026vei,Ando:2026poq}, especially for images with negative parity~\cite{Ando2026SaddleImages}. Stochastic diffraction can therefore open a new window onto dark-matter theories beyond CDM~\cite{Thakore:2026waw,Nadler:2026waz} and onto non-standard initial conditions imprinted on small-scale halos~\cite{Bringmann:2025cht}.

ULDM is an extreme realization of this picture. Rather than hosting a population of discrete lenses, a ULDM halo is itself a diffractive barrier: its granules form a random screen of time delays with order-unity fluctuations and a single coherence length $\ell_c$, set by the de Broglie wavelength. Since the granules rearrange at nonrelativistic speeds, they are effectively frozen while the wave crosses them, so this diffraction by a static screen is distinct from the modulation induced by the oscillation of the field\footnote{The power spectrum of the gravitational potentials is multi-structured: at frequencies $f\simeq m_\phi/(\pi\hbar)$, the scalar field oscillations source a rapidly oscillating component. Its contribution to lensing is suppressed because the oscillating gravitational potentials are smaller by order $\sigma_v^2$ than the slowly varying potentials sourced by the granules, and their rapid oscillations largely average out along the propagation path through the halo~\cite{KhmelnitskyRubakov2014,Kim2024Astrometry}. The lower frequency component, at $f\lesssim m_\phi\sigma_v^2/(2\pi\hbar)$, describes the slow evolution of the granule pattern considered here~\cite{KimMitridate2023PTA}.}. The situation resembles X-ray crystallography, in which a crystal diffracts radiation whose wavelength is comparable to its lattice spacing (here the Fresnel scale plays the role of the wavelength). Unlike a crystal, however, the granules are not arranged on a lattice: instead of sharp Bragg peaks, the screen produces a random interference pattern whose statistics encode the two-point function of the field, as in diffuse scattering by liquids and amorphous solids~\cite{Guinier:1963}. A chirping binary sweeps its Fresnel scale across the screen, $r_F\propto f^{-1/2}$, so a single GW event scans the granules from the averaged regime at low frequencies, through maximal diffraction at $r_F\sim\ell_c$, towards geometric optics at high frequencies.

Because the granules fill the lens halo, every strongly lensed image is seen through such a screen, precisely where the host magnification enhances its effect. Previous studies of GW lensing by ULDM have captured parts of this picture. Solitonic cores have been treated as smooth single lenses, both for microlensing of signals in future detectors~\cite{Fairbairn:2022xln} and for single-image signatures in LISA~\cite{Singh2025LISA_FDM}. The granules themselves have so far been treated perturbatively: Ref.~\cite{OguriTakahashi2022AmpPhase} computed, at Born order, the amplitude and phase dispersions they induce around strongly magnified images and their correlations between frequencies, without obtaining the wave-optics amplification factor across the band. Wave-optics distortions of strongly lensed LISA signals by ULDM, NFW and self-interacting substructure have also been compared~\cite{Liu2026WaveOpticsSubstructure}. Saddle images, common in strong lensing, carry a distinct Morse phase and are harder to treat in wave optics~\cite{Dai:2017huk,Ando2026SaddleImages}. 
A full wave-optics characterization of the ULDM imprint, including non-perturbative effects, a consistent realization of the granular field and the impact of the macro-potential, therefore remains to be performed.

In this work, we compute the full wave-optics amplification factor of a strongly lensed GW that passes through the granular field of its lens halo (Fig.~\ref{fig:overview-sis}). We describe ULDM by a complex Gaussian random field and derive an approximate probability distribution for its projected surface density, quantifying how departures from Gaussianity depend on the mean halo profile and coherence length. For the lensing calculations, we numerically generate  surface density realizations directly from the field covariance through a Karhunen--Lo\`eve expansion along the line of sight (Sec.~\ref{sec:uldm-stats}). We take the host to be a singular isothermal sphere, convert the granule maps into lensing potentials, and combine them with the convergence and shear of the host, which set the parity of the macroimage (Sec.~\ref{sec:methods}). After reviewing wave-optics lensing, we evaluate $F(w)$ with a Fourier-space form of the diffraction integral that applies to minima and saddles alike (Sec.~\ref{sec:gwlensing}), and validate it against the contour integration of \texttt{GLoW}~\cite{VillarrubiaRojo2024GLoW}, which we extend to lenses defined by an arbitrary tabulated density field (Appendix~\ref{app:glow}). The method is general: granule maps from wave simulations, populations of CDM subhalos or other host profiles can be used without modification, provided the map is periodic on the box on which it is defined.

We consider sightlines ranging from weakly lensed regions far outside the Einstein radius of the lens halo, through strongly magnified minima near its critical curve, to saddles inside it, which can be demagnified. For each sightline, we assess the detectability of the granular signature in LISA observations of massive black hole binaries (Sec.~\ref{sec:detectability}). The granules imprint percent-level, frequency-dependent fluctuations once the Fresnel scale reaches their coherence length. These appear in the \emph{amplification factor} $F(f)$ (cf. Eq.~\eqref{eq:hlensed}), which modifies the original waveform, and is the key observable we consider. 
These fluctuations are enhanced to $\gtrsim10\%$ near the critical curve and for saddle images, which at matched magnification lie closer to the halo centre and therefore see a stronger granule field. For a $10^6\,M_\odot$ binary, the signature is distinguishable from an unlensed waveform over $m_\phi\sim10^{-22}$--$5\times10^{-21}$~eV.

We discuss these results and conclude in Sec.~\ref{sec:conclusions}, leaving ground-based frequencies, which probe heavier ULDM, to future work. We use units with $c=1$, retaining $G$ and $\hbar$ explicitly. Numerical values are quoted in conventional astrophysical units.

%%%%%
\section{Ultralight Dark Matter and Density Fluctuations}
\label{sec:uldm-stats}
%%%%%

Our lensing calculation uses surface density realizations constructed from a complex Gaussian ULDM field $\Psi$. This connects the projected fluctuations to the statistics of the underlying field and does not force their distribution to be Gaussian. Sections~\ref{sec:uldm-stats:picture} and~\ref{sec:uldm-stats:covariance} introduce the halo model and its three-dimensional density statistics. Section~\ref{sec:uldm-stats:surface} develops a coherent patch approximation to understand the projected distribution and its departure from Gaussianity. Section~\ref{sec:uldm-stats:kl} describes the numerical construction of surface density realizations, and Sec.~\ref{sec:uldm-stats:sis} specializes the results to our singular isothermal sphere (SIS) benchmarks.

\begin{figure*}[t]
\centering
\includegraphics[width=0.65\textwidth]{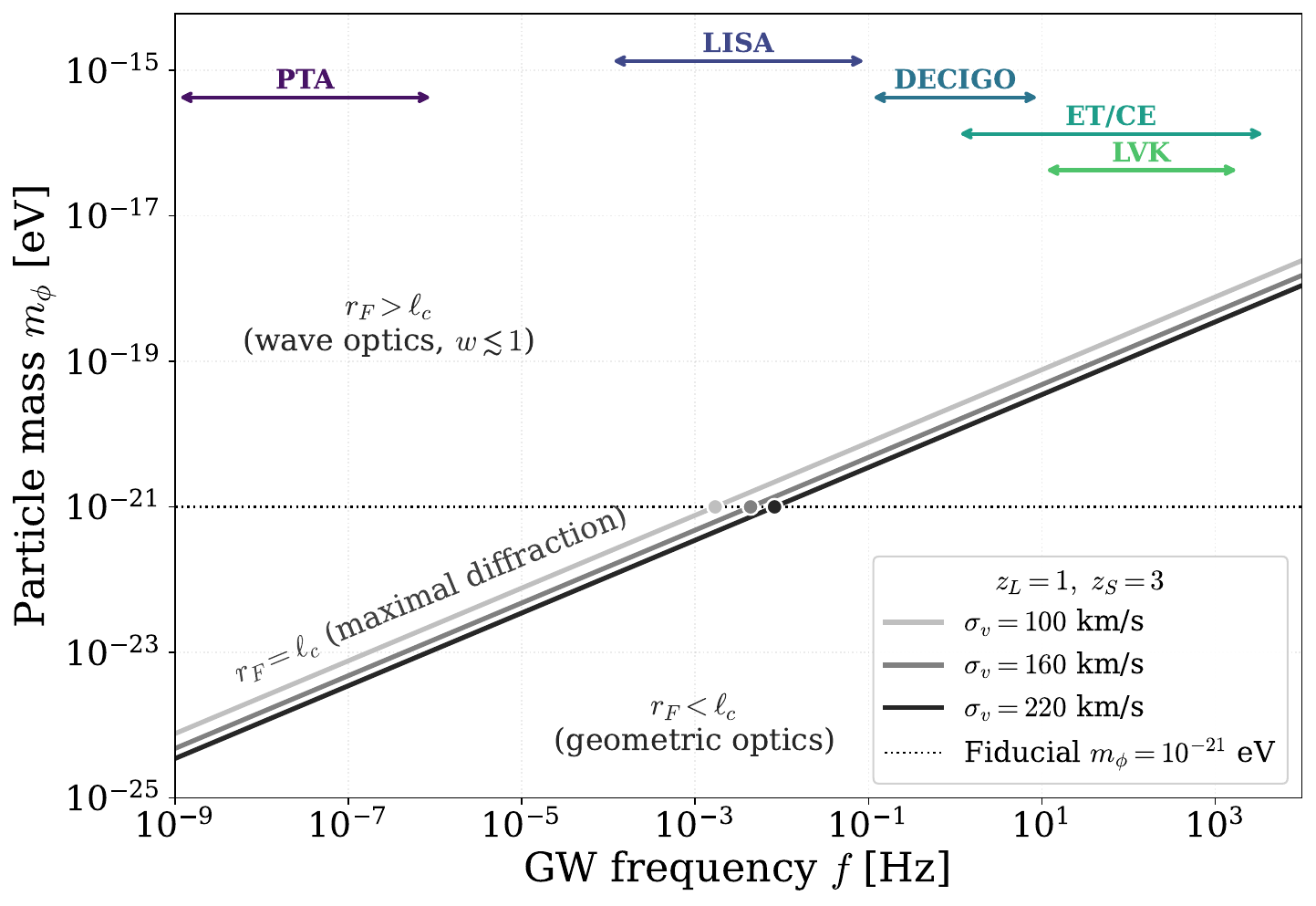}
\caption{Wave and geometric optics regimes over GW frequency and ULDM particle mass, for lens and source redshifts $z_L=1$ and $z_S=3$ (used throughout). Solid lines show $r_F(f)=\ell_c(m_\phi,\sigma_v)$, i.e.\ $w\equiv(\ell_c/r_F)^2=1$ [Eq.~\eqref{eq:w-fresnel}], for each $\sigma_v$ in Table~\ref{tab:physics-cases}. Wave effects dominate in the upper left ($w<1$), while the lower right ($w>1$) approaches geometric optics for $w\gg1$. Gray dots indicate the fiducial $m_\phi=10^{-21}$~eV. The LISA band brackets all three curves near this mass, while PTA and ground-based detectors (LIGO/Virgo, ET/CE) would require lighter and heavier $m_\phi$, respectively.}
\label{fig:param-space}
\end{figure*}

\subsection{Physical picture and statistical assumptions} \label{sec:uldm-stats:picture}

ULDM density granules arise from interference between waves with different velocities. Simulations of self-gravitating ULDM halos find a central solitonic core surrounded by a fluctuating envelope whose mean density is approximately NFW-like~\cite{Schive:2014dra,Schive:2014hza,Mocz:2017wlg,Veltmaat:2018dfz}. The core is a coherent, gravitationally supported structure. In the surrounding halo, many waves with different phases overlap, producing transient density granules. We focus on this granular region outside the core and write $\rho=\rho_{\rm sm}+\delta\rho$, where $\rho_{\rm sm}\equiv\langle\rho\rangle$ is the ensemble mean ULDM halo density. The subscript $\mathrm{sm}$ denotes this smooth component.

For a real scalar with negligible nongravitational self interactions, we define its nonrelativistic envelope by
\begin{equation}
\phi(\bm x,t)=\frac{1}{\sqrt{2}}\left[
e^{-ik_\phi t}\Psi(\bm x,t)+\mathrm{c.c.}\right],
\qquad k_\phi=\frac{m_\phi}{\hbar}.
\label{eq:uldm-envelope}
\end{equation} 
The complex envelope $\Psi$ records the amplitude and phase of the real field and varies on timescales much longer than $k_\phi^{-1}$. Averaging over the carrier period gives $\rho=k_\phi^2|\Psi|^2$ to leading nonrelativistic order.

The field $\Psi$ evolves according to the coupled Schr\"odinger and Poisson equations~\cite{HuBarkanaGruzinov2000,HuiOstrikerTremaineWitten2017}. We approximate $\Psi$ locally by free Schr\"odinger waves~\cite{Bar-Or:2018pxz,Hui:2020hbq}. In a locally freely falling frame, we assume that the smooth halo's tidal forces and the self gravity of the granules produce relative displacements much smaller than the coherence length over one coherence time. The mean density $\rho_{\rm sm}$ is taken to vary slowly across the coherence length $\ell_c$ introduced below, $\ell_c|\nabla\ln\rho_{\rm sm}|\sim \ell_c/R_h\ll1$, with $R_h$ the radius of the halo. We also approximate the velocity dispersion as constant over the region contributing appreciably to the surface density. These assumptions apply to the granular halo outside the coherent core.

The superposition of many waves with independent random phases motivates a complex Gaussian random field description of $\Psi$ in this region~\cite{Centers:2019dyn,Hui:2020hbq,Kim2024Astrometry,Cheong:2024ose}. At any fixed position, its real and imaginary parts are independent Gaussian variables with zero mean and equal variance. The Fourier ensemble and its relation to the velocity distribution are specified in Appendix~\ref{app:uldm-wave}.

At a fixed point, the squared modulus of this amplitude gives the density distribution~\cite{Centers:2019dyn,Chan2020FDMCriticalCurves}
\begin{equation}
p(\rho)=\frac{1}{\rho_0}e^{-\rho/\rho_0},
\qquad \rho\geq0,\qquad
\rho_0\equiv\langle\rho\rangle.
\label{eq:rho-exponential}
\end{equation}
The exponential distribution has variance $\langle(\delta\rho)^2\rangle=\rho_0^2$, so the three dimensional density contrast is of order unity. This result holds for general velocity spectra within the ensemble of Eq.~\eqref{eq:uldm-mode-pdf}, including the Maxwellian distribution adopted below.

\subsection{Covariance and coherence scales}
\label{sec:uldm-stats:covariance}

The velocity distribution determines the spatial and temporal correlations of the granules. We adopt an isotropic Maxwellian velocity distribution with one-dimensional dispersion $\sigma_v$ and zero mean velocity in the halo rest frame. For a locally homogeneous region with mean density $\rho_0$, and for frequencies $f\ll m_\phi/(2\pi\hbar)$, Wick's theorem gives the density covariance~\cite{Bar-Or:2018pxz,Hui:2020hbq,Kim2023Interferometers}
\begin{equation}
C_{\delta\rho}(\tau,r)=\rho_0^2 \left(1+\frac{\tau^2}{\tau_c^2}\right)^{-3/2} \exp\left[-\frac{r^2/\ell_c^2}{1+\tau^2/\tau_c^2}\right].
\label{eq:drho-2pt}
\end{equation}
Here $C_{\delta\rho}(\tau,r)=\langle\delta\rho(t,\bm x)\delta\rho(t',\bm x')\rangle$, with $\tau=t-t'$ and $r=|\bm x-\bm x'|$. 
The characteristic spatial and temporal scales are
\begin{equation}
\ell_c=\frac{\hbar}{m_\phi\sigma_v} =19.2\,\mathrm{pc} \left(\frac{10^{-21}\,\mathrm{eV}}{m_\phi}\right) \left(\frac{100\,\mathrm{km/s}}{\sigma_v}\right),
\label{eq:coherence-length}
\end{equation}
\begin{equation}
\tau_c=\frac{\ell_c}{\sigma_v} =1.88\times10^5\,\mathrm{yr} \left(\frac{10^{-21}\,\mathrm{eV}}{m_\phi}\right) \left(\frac{100\,\mathrm{km/s}}{\sigma_v}\right)^2.
\label{eq:coherence-time}
\end{equation}
Our $\ell_c$ is the reduced de Broglie scale for the 1-D dispersion, equal to $l_c/(2\pi)$. At equal time, $C_{\delta\rho}(0,r)=\rho_0^2e^{-r^2/\ell_c^2}$. In the lensing calculation, we adopt a static lens and use this equal-time covariance.\footnote{A sufficient condition is that $T_{\rm obs}/(1+z_L)\ll \tau_c$. The other relevant condition is that the light-crossing time of each over- or under-density (which are the time-dependent structures in the problem) is much shorter than $\tau_c$.  This is always satisfied from Eq.~\eqref{eq:coherence-length}, and the fact that $\sigma_v\ll c$. Hence, cells at different depths are sampled at different times and are statistically independent. As a result, the projected statistics coincide with the equal-times one.
}

The large density contrast motivates searches for granule-induced lensing perturbations, including image and flux ratio anomalies~\cite{Chan2020FDMCriticalCurves,Laroche:2022pjm,Amruth2023EinsteinRings}, time delay anomalies~\cite{Liu:2024uvr}, perturbations of extended radio arcs~\cite{Powell2023VLBI_FDM}, cluster microlensing signatures~\cite{Broadhurst:2024ggk}, and GW amplitude and phase fluctuations~\cite{OguriTakahashi2022AmpPhase}. Figure~\ref{fig:param-space} relates $\ell_c$ to the GW frequencies at which diffraction becomes relevant. The lensing perturbations depend on the projected surface density, so we next examine how the projection changes the granule statistics.

\subsection{Surface density statistics}
\label{sec:uldm-stats:surface}

Lensing responds to the projected surface density, whose fluctuations depend on both the three dimensional density contrast and the distribution of matter along the line of sight. We obtain the projected statistics from the complex Gaussian field $\Psi$ and examine how the probability distribution of the surface density differs from the exponential density distribution at a point in three dimensions.
Several lensing studies approximate the projected fluctuations as a Gaussian random field, motivated by the central limit theorem~\cite{Chan2020FDMCriticalCurves,Powell2023VLBI_FDM,Amruth2023EinsteinRings,Liu:2024uvr}. The dependence of departures from Gaussianity on the density profile and coherence length calls for a quantitative description. Other descriptions use randomly placed three dimensional Gaussian density clumps~\cite{Kawai:2021znq}, or two dimensional Gaussian profiles with signed amplitudes calibrated against numerical halo realizations~\cite{Laroche:2022pjm}.
Our model also remains an approximation to a dynamical halo, whose physical accuracy must ultimately be assessed against simulations~\cite{Schive:2014dra,Mocz:2017wlg,Veltmaat:2018dfz,Chan2020FDMCriticalCurves}.

We first use a coherent patch approximation to obtain the probability distribution of the surface density at a fixed position in the lens plane. This approximation treats the surface density as a sum of independent contributions from regions of depth $\sim\ell_c$ along the corresponding line of sight. We use the resulting distribution to identify when the surface density distribution departs appreciably from a Gaussian. For lensing, we will construct surface density realizations numerically by projecting the three-dimensional density $\rho=k_\phi^2|\Psi|^2$, retaining the spatial correlations implied by the complex Gaussian random field $\Psi$. In Sec.~\ref{sec:uldm-stats:kl}, we will evaluate this projection using a Karhunen--Lo\`eve (KL) expansion, which expresses the surface density in terms of two-dimensional complex Gaussian fields.

With $z$ the line-of-sight coordinate and $\bm\xi$ the position in the lens plane, we define the projected ULDM surface density $\Sigma$ and its smooth component $\Sigma_{\rm sm}$ by
\begin{equation}
\begin{split}
\Sigma(\bm\xi)&=\int dz\,\rho(\bm\xi,z),\\
\Sigma_{\rm sm}(\bm\xi)&=\int dz\,\rho_{\rm sm}(\bm\xi,z),\qquad \delta\Sigma=\Sigma-\Sigma_{\rm sm}.
\end{split}
\label{eq:surface-density-definition}
\end{equation}
Under the static approximation, we use the equal-time covariance of $\Psi$, which determines the spatial correlations of the granule density fluctuations around the smooth halo profile. For the Maxwellian velocity distribution with constant $\sigma_v$, a region with constant mean density $\rho_0$ has $C_\Psi(0,r)=(\rho_0/k_\phi^2)e^{-r^2/(2\ell_c^2)}$. We extend this expression to a mean halo density $\rho_{\rm sm}$ varying slowly over $\ell_c$,
\begin{equation}
C_\Psi(\bm x,\bm x')=\frac{\sqrt{\rho_{\rm sm}(\bm x)\rho_{\rm sm}(\bm x')}}{k_\phi^2}e^{-|\bm x-\bm x'|^2/(2\ell_c^2)}.
\label{eq:psi-inhomogeneous}
\end{equation}
Wick's theorem and line-of-sight integration give the projected covariance derived in Appendix~\ref{app:uldm-projection}. When $\rho_{\rm sm}$ varies little over $\ell_c$ along the line of sight, the variance reduces to~\cite{Chan2020FDMCriticalCurves,Powell2023VLBI_FDM}
\begin{equation}
\operatorname{Var}\Sigma(\bm\xi)\simeq\sqrt{\pi}\ell_c\int dz\,\rho_{\rm sm}^2(\bm\xi,z).
\label{eq:surface-variance-local}
\end{equation}
A useful measure of the fractional fluctuation is
\begin{equation}
\begin{split}
N_{\rm eff}(\bm\xi)&\equiv\frac{\Sigma_{\rm sm}^2(\bm\xi)}{\operatorname{Var}\Sigma(\bm\xi)}\simeq\frac{L_{\rm eff}(\bm\xi)}{\sqrt{\pi}\ell_c},\\
L_{\rm eff}(\bm\xi)&\equiv\frac{\left[\int dz\,\rho_{\rm sm}(\bm\xi,z)\right]^2}{\int dz\,\rho_{\rm sm}^2(\bm\xi,z)}.
\end{split}
\label{eq:neff}
\end{equation}
Here $L_{\rm eff}$ is the effective line-of-sight depth determined by the mean density profile, also introduced by Kawai et al.~\cite{Kawai:2021znq}. The quantity $N_{\rm eff}$ gives the fractional variance, $\langle(\delta\Sigma/\Sigma_{\rm sm})^2\rangle=N_{\rm eff}^{-1}$.

Within the small region of the lens plane considered here, we neglect transverse variations of the mean density profile. The projected covariance and power spectrum are
\begin{equation}
\begin{split}
C_{\delta\Sigma}(R)&=\operatorname{Var}\Sigma\,e^{-R^2/\ell_c^2},\\
P_\delta(k_\perp)&=\frac{\pi\ell_c^2}{N_{\rm eff}}e^{-k_\perp^2\ell_c^2/4},\qquad \delta\equiv\frac{\delta\Sigma}{\Sigma_{\rm sm}},
\end{split}
\label{eq:projected-power}
\end{equation}
where $R$ is transverse separation and $C_\delta(\bm R)=\int d^2k_\perp\,P_\delta(k_\perp)e^{i\bm k_\perp\cdot\bm R}/(2\pi)^2$. The same dependence on the coherence scale and profile-weighted depth appears in earlier descriptions of the projected fluctuations~\cite{Chan2020FDMCriticalCurves,Kawai:2021znq}. Projection reduces the fractional fluctuation while retaining the transverse correlation scale.

To estimate the one-point distribution, we adopt a coherent patch approximation in which densities in different line-of-sight patches are treated as independent. Each patch has mean density $\rho_{\rm sm}$ at its position. The density covariance decays over $\ell_c$, and its normalized integral along the line of sight defines the effective patch depth $\Delta z_{\rm eff}=\sqrt{\pi}\ell_c$. This choice makes the independent patch approximation reproduce the leading projected variance in Eq.~\eqref{eq:surface-variance-local}. The subscript $\mathrm{patch}$ denotes quantities calculated in the independent patch approximation, including probability densities, ensemble averages, and variances. The patch construction and its Fourier transform are derived in Appendix~\ref{app:uldm-projection}.

The joint probability density of the total surface density is the product of the exponential distributions in Eq.~\eqref{eq:rho-exponential} for the individual patches. For a slowly varying mean profile, the sum of their surface density contributions has the distribution (cf. Eq.~\eqref{eq:patch-pdf-fourier})
\begin{equation}
\begin{split}
p_{\rm patch}(\Sigma|\bm\xi)\simeq{}&\int_{-\infty}^{\infty}\frac{ds}{2\pi}\exp\Bigg\{is\Sigma\\
&-\frac{1}{\Delta z_{\rm eff}}\int dz\,\ln\left[1+is\Delta z_{\rm eff}\rho_{\rm sm}(\bm\xi,z)\right]\Bigg\}.
\end{split}
\label{eq:patch-pdf-general}
\end{equation}
Here $s$ is the Fourier variable conjugate to $\Sigma$. In other words, this is the one-point PDF of the fluctuating $\Sigma$ at a fixed position $\bm \xi$.

The Gaussian limit follows by keeping the logarithm through quadratic order,
\begin{equation}
\begin{split}
\ln(1+is\Delta z_{\rm eff}\rho_{\rm sm})
&\simeq is\Delta z_{\rm eff}\rho_{\rm sm}\\
&\quad+\frac{s^2(\Delta z_{\rm eff})^2\rho_{\rm sm}^2}{2}.
\end{split}
\label{eq:patch-log-quadratic}
\end{equation}
The ratio of the surface density standard deviation from one coherent patch to that of the total surface density is approximately $(L_{\rm eff}\rho_{\rm sm}/\Sigma_{\rm sm})/\sqrt{N_{\rm eff}}$, using Eqs.~\eqref{eq:rho-exponential} and~\eqref{eq:neff}. The quadratic approximation is justified when this ratio is small throughout the line of sight. Substituting this expansion into Eq.~\eqref{eq:patch-pdf-general} and carrying out the Fourier integral gives
\begin{equation}
\begin{split}
&p_{\rm patch}(\Sigma|\bm\xi)\simeq
\sqrt{\frac{N_{\rm eff}}{2\pi\Sigma_{\rm sm}^2}}\\
&\qquad\times\exp\left[-\frac{N_{\rm eff}}{2}
\left(\frac{\Sigma-\Sigma_{\rm sm}}{\Sigma_{\rm sm}}\right)^2\right].
\end{split}
\label{eq:patch-gaussian-limit}
\end{equation}
The mean is $\Sigma_{\rm sm}$ and the variance is $\Sigma_{\rm sm}^2/N_{\rm eff}$.

Departures from a Gaussian can be quantified by skewness, which measures asymmetry, and excess kurtosis, which is sensitive to the tails relative to a Gaussian. Both vanish for a Gaussian distribution. Equation~\eqref{eq:patch-pdf-general} gives a dependence on $N_{\rm eff}^{-1/2}$ and $N_{\rm eff}^{-1}$, respectively, with additional factors determined by the mean density profile. These quantities can also be calculated directly from the covariance of the complex Gaussian field $\Psi$. For the covariance in Eq.~\eqref{eq:psi-inhomogeneous}, the two calculations give the same $N_{\rm eff}$ and profile dependence at leading order in the slowly varying approximation, with numerical coefficients differing by factors of order unity. For example, retaining the continuous correlations of $\Psi$ increases the skewness by a factor $2/\sqrt{3}$ relative to the coherent patch result. The profile integrals and this comparison are derived in Appendix~\ref{app:uldm-projection}.

The coherent patch approximation relates these trends to the sum of independent surface density contributions along the line of sight and provides the distribution itself through Eq.~\eqref{eq:patch-pdf-general}. For the lensing calculation, we generate surface density realizations from the covariance of $\Psi$ using the method in Sec.~\ref{sec:uldm-stats:kl}.

\subsection{Karhunen--Lo\`eve construction}
\label{sec:uldm-stats:kl}

Our aim is to generate nonnegative total surface-density realizations with the statistics implied by the complex Gaussian random field $\Psi$. We use a Karhunen--Lo\`eve (KL) expansion~\cite{GhanemSpanos1991} to construct these realizations from two dimensional fields. This also avoids the memory cost of storing a full three dimensional field. We first expand $\Psi$ in eigenfunctions of its line-of-sight covariance,
\begin{equation}
\Psi(\bm\xi,z)=\frac{1}{k_\phi}\sum_n\sqrt{\lambda_n}\,u_n(z)\eta_n(\bm\xi).
\label{eq:kl-amplitude}
\end{equation}
Here $u_n(z)$ form an orthonormal basis along the line of sight, and $\eta_n(\bm\xi)$ are independent two dimensional complex Gaussian random fields with zero mean and unit variance. The $\lambda_n\ge0$ are the eigenvalues of the density weighted longitudinal kernel. Squaring this expansion produces cross terms containing $u_n(z)u_m^*(z)$ for $n\ne m$. These terms vanish upon line-of-sight integration because $\int dz\,u_n(z)u_m^*(z)=\delta_{nm}$, giving
\begin{equation}
\Sigma(\bm\xi)\equiv k_\phi^2\int dz\,|\Psi(\bm\xi,z)|^2=\sum_n\lambda_n|\eta_n(\bm\xi)|^2.
\label{eq:kl-sum}
\end{equation}
We therefore generate the independent $\eta_n$ fields with the transverse covariance inherited from $\Psi$ and accumulate their weighted squared amplitudes. This construction uses the covariance in Eq.~\eqref{eq:psi-inhomogeneous} and the same local approximation as Eq.~\eqref{eq:projected-power}, in which neighboring lines of sight share a common mean density profile along $z$. Appendix~\ref{app:uldm-kl} specifies the kernel, transverse covariance, and numerical evaluation.

The computational cost of this construction is largely determined by the evaluation of the KL weights $\lambda_n$ in Eq.~\eqref{eq:kl-sum}, which are the eigenvalues of the density-weighted longitudinal kernel. On $n$ line-of-sight quadrature nodes (Appendix~\ref{app:uldm-kl}) this kernel becomes an $n\times n$ matrix, and a dense symmetric eigensolver needs $\mathcal{O}(n^{3})$ operations. On two Intel Xeon Platinum 8358 CPUs with 128 threads, the eigendecomposition takes about $2.4$~min for $n=1.7\times10^{4}$ and about $20$~min for $n=3.4\times10^{4}$, consistent with $n^3$ scaling, while on a single NVIDIA A100 GPU $n=3.4\times10^{4}$ takes about $4$~min. Because the weights depend only on $(\sigma_v,m_\phi,\xi,n)$, they are computed once and reused for every realization. Drawing a map then takes one inverse FFT per retained mode on the $N^{2}$ grid, and nearly all $n$ modes are retained, so a map costs $\mathcal{O}(nN^{2}\log N)$.

\subsection{SIS lens model and position-dependent granule statistics}
\label{sec:uldm-stats:sis}

We adopt a singular isothermal sphere (SIS) as the smooth lens halo (i.e.~macrolens) throughout our lensing calculations. This simple model is specified by a single constant one-dimensional velocity dispersion $\sigma_v$, with density and projected surface density
\begin{equation}
\rho_{\rm sm}(r)=\frac{\sigma_v^2}{2\pi G r^2},\qquad \Sigma_{\rm sm}(\xi)=\frac{\sigma_v^2}{2G\xi}.
\label{eq:sis-host-profile}
\end{equation}
Here $r^2=\xi^2+z^2$, and $\xi=|\bm\xi|$ is the projected distance from the halo center. In the lensing calculation, we evaluate the profile at an unperturbed macroimage position $\bm\xi_0$ of the smooth host and generate surface density realizations in its neighborhood. The smooth host supplies the local convergence and shear, while $\delta\Sigma$ supplies the granule perturbation in Sec.~\ref{sec:methods}.

For the SIS, integration of the mean profile and its square gives $L_{\rm eff}=2\pi\xi$ and hence
\begin{equation}
\begin{split}
N_{\rm eff}(\xi)&\simeq\frac{2\sqrt{\pi}\xi}{\ell_c}\\
&\simeq740\left(\frac{\xi}{4\,\mathrm{kpc}}\right)\left(\frac{m_\phi}{10^{-21}\,\mathrm{eV}}\right)\left(\frac{\sigma_v}{100\,\mathrm{km/s}}\right).
\end{split}
\label{eq:neff-sis}
\end{equation}
Closer to the halo center, most of the mean surface density comes from a shorter interval along the line of sight, as reflected in $L_{\rm eff}=2\pi\xi$. At fixed coherence length $\ell_c$, fewer coherent patches then contribute appreciably to the surface density. This reduces $N_{\rm eff}$ and increases the fractional standard deviation, $\sqrt{\operatorname{Var}\Sigma}/\Sigma_{\rm sm}=N_{\rm eff}^{-1/2}$.

For the SIS, the sufficient condition for the Gaussian limit reduces to $2/\sqrt{N_{\rm eff}}\ll1$. The coherent patch estimates of the skewness and excess kurtosis are
\begin{equation}
\begin{split}
\frac{\langle(\delta\Sigma)^3\rangle_{\rm patch}}{\left(\operatorname{Var}_{\rm patch}\Sigma\right)^{3/2}}&\simeq\frac{3}{\sqrt{N_{\rm eff}}},\\
\frac{\langle(\delta\Sigma)^4\rangle_{\rm patch}}{\left(\operatorname{Var}_{\rm patch}\Sigma\right)^2}-3&\simeq\frac{15}{N_{\rm eff}}.
\end{split}
\label{eq:sis-patch-nongaussian}
\end{equation}
Within the regime $\ell_c\ll\xi$ outside the solitonic core, moving closer to the halo center reduces $N_{\rm eff}$ and increases the skewness of the surface density distribution. For illustration, $\xi=4$~kpc, $\sigma_v=100$~km/s, and $m_\phi=10^{-22}$~eV give $\ell_c=192$~pc and $N_{\rm eff}\simeq74$. The coherent patch approximation then predicts a skewness of about $0.35$ and an excess kurtosis of about $0.20$; retaining the continuous correlations of $\Psi$ raises these to about $0.40$ and $0.29$ (Appendix~\ref{app:uldm-projection}).

The SIS model extends to infinite radius and has an unbounded total mass. Its $L_{\rm eff}$ grows linearly with $\xi$. A finite halo boundary can instead shorten the effective line-of-sight depth near the outskirts, as illustrated for a truncated NFW halo by Kawai et al.~\cite{Kawai:2021znq}. This can reduce $N_{\rm eff}$ and increase departures from Gaussianity.

Table~\ref{tab:physics-cases} summarizes the fiducial halo parameters at $\xi=4$~kpc and $m_\phi=10^{-21}$~eV. The tabulated $N_{\rm eff}$ values imply modest departures from Gaussianity according to the coherent patch estimates in Eq.~\eqref{eq:sis-patch-nongaussian}. Halos~A and B are the Type~I examples of Fig.~\ref{fig:amplification}, and Halo~C is the benchmark host of Sec.~\ref{sec:detectability}.

%%%%%
\begin{table}[t]
\centering
\begin{ruledtabular}
\begin{tabular}{lccc}
 & Halo A & Halo B & Halo C \\
\colrule
$\sigma_v$ [km/s]          & 100 & 160 & 220 \\
$\ell_c$ [pc]              & 19.2 & 12.0 & 8.7 \\
$N_{\rm eff}$              & 739 & 1182 & 1625 \\
$L_{\rm box}$ [pc]         & 208.2 & 164.6 & 140.3 \\
$L_{\rm box}/\ell_c$       & 10.8 & 13.7 & 16.1 \\
$\Ml(<\xi)$ [$M_\odot$]      & $1.9\times10^{10}$ & $4.8\times10^{10}$ & $9.0\times10^{10}$ \\
$M_{\rm eff}$ [$M_\odot$] & $2.4\times10^{6}$ & $9.3\times10^{5}$ & $4.9\times10^{5}$ \\
$f(w=1)$ [mHz] & $1.7$ & $4.4$ & $8.2$ \\
\end{tabular}
\caption{ULDM lens halo benchmarks at fixed $m_\phi=10^{-21}$~eV and $\xi=4$~kpc, for $z_L=1$ and $z_S=3$. $\Ml(<\xi)$ is the SIS mass within a sphere of radius $\xi$, $M_{\rm eff}=\pi\Sigma_{\rm crit}\ell_c^{2}$ is the lens mass whose Einstein radius equals $\ell_c$, and $f(w=1)$ is the GW frequency at which the Fresnel scale equals $\ell_c$ (Sec.~\ref{sec:gwlensing}). The values of $N_{\rm eff}$ are the analytic SIS predictions at the stated position.}
\label{tab:physics-cases}
\end{ruledtabular}
\end{table}
%%%%%

%%%%%
\section{Lens model}
\label{sec:methods}
%%%%%
We construct a lens model in which the mean profile of the lens halo produces a macroimage and the ULDM granules add a fluctuating surface density around it. The macroimage is characterized by the local convergence and shear of the host, while the granules produce frequency-dependent amplitude and phase modulations through diffraction. We use a local macrolens setup similar to that adopted for populations of stellar point masses~\cite{Diego:2019lcd,Mishra:2021xzz,Shan:2023qvd,Shan:2026beq} and dark matter subhalos~\cite{Ando:2026poq,Ando2026SaddleImages}, with ULDM granules supplying the perturbing density field in our calculation. We generate the surface density realizations as described in Sec.~\ref{sec:uldm-stats:kl} and calculate the amplification factor $F(w)$ for the signal associated with a single macroimage.

\begin{figure*}[tbp]
\centering
\includegraphics[width=\textwidth]{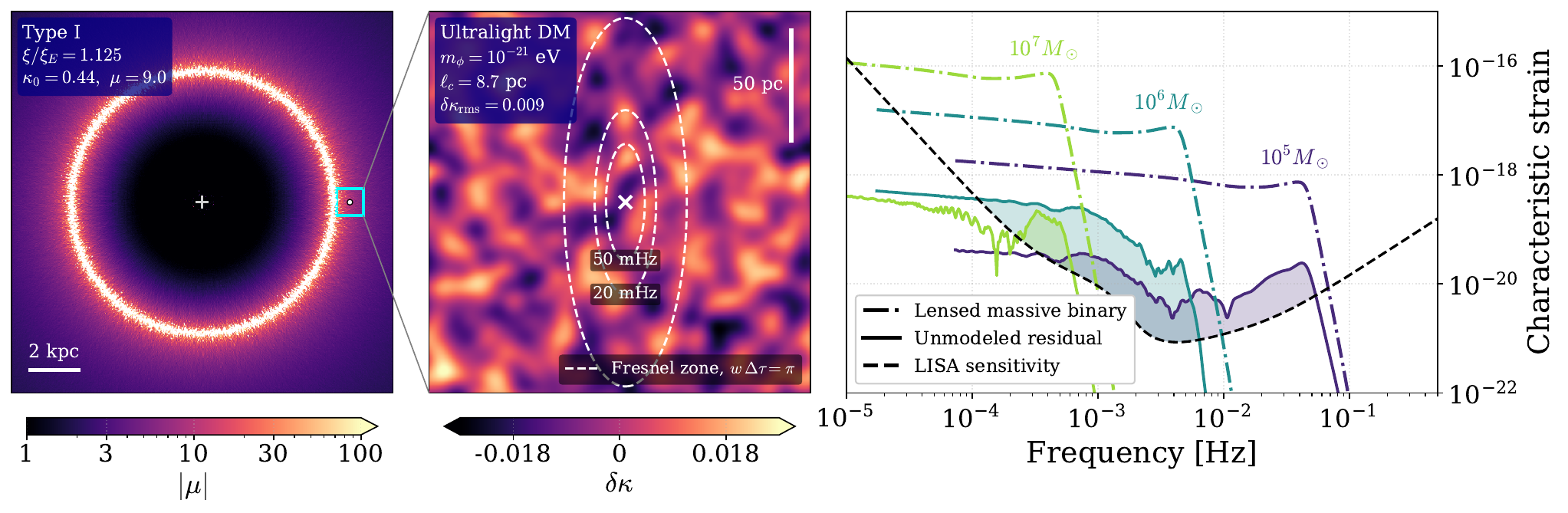}\\[2pt]
\includegraphics[width=\textwidth]{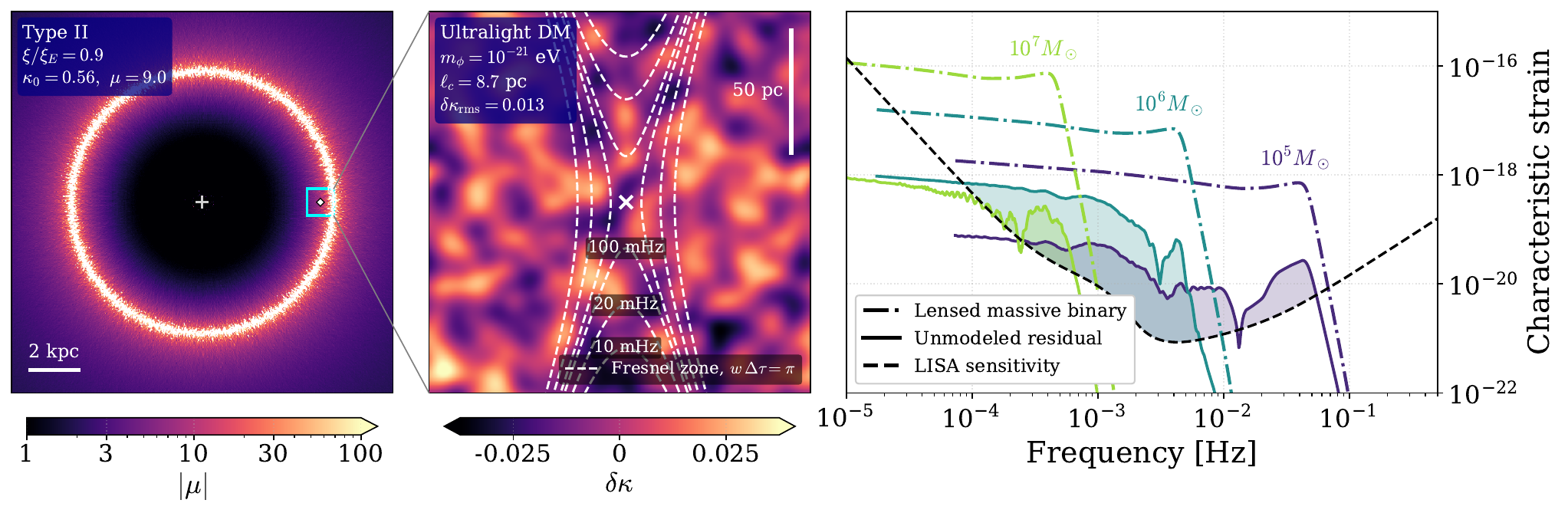}
\caption{The imprint of ULDM granules on a strongly lensed gravitational wave observed by LISA, for a type~I (minimum) and a type~II (saddle) macroimage. The two rows have different source positions with the same smooth host magnification, $|\mu_0|=9$. The host is an SIS with $\sigma_v=220$~km/s, and $m_\phi=10^{-21}$~eV ($\ell_c=8.7$~pc), $z_L=1$, $z_S=3$. \emph{\textbf{Top.}} Type~I ($\xi/\xi_E=1.125$, $\kappa_0=0.44$). \emph{\textbf{Bottom.}} Type~II ($\xi/\xi_E=0.9$, $\kappa_0=0.56$). \emph{\textbf{Left}}: $|\mu|$ of the SIS host with granules, and its critical curve, the box marks the image (not to scale). \emph{\textbf{Centre}}: granule convergence $\delta\kappa$ around the image, with Fresnel zones $w\,\Delta\tau=\pi$ (dashed). \emph{\textbf{Right}}: lensed strain (dash-dotted) of the three MBHB benchmarks, modelled with \textsc{IMRPhenomXAS}, residual after refitting amplitude, phase and time (solid), and LISA noise (dashed). Both rows use the same granule realization.}
\label{fig:overview-sis}
\end{figure*}

Figure~\ref{fig:overview-sis} illustrates the physical setup. The granules perturb the magnification map across the lens plane, including the neighborhood of each macroimage. The granules modify the propagation phase and produce frequency-dependent diffraction patterns when the Fresnel scale $r_F$ (Sec.~\ref{sec:gwlensing}) is comparable to $\ell_c$.\footnote{At very high macroimage magnification, the effective Fresnel scale can substantially exceed $r_F$~\cite{OguriTakahashi2022AmpPhase}, as illustrated by the elongated Fresnel zones in Fig.~\ref{fig:overview-sis}.} The perturbations to the magnification map are especially pronounced near the host critical curve. The resulting waveform can retain a frequency-dependent residual after refitting its overall amplitude, phase and arrival time. We specify the local lens model and its SIS realization in Secs.~\ref{sec:methods:fw} and~\ref{sec:methods:sis}; Sec.~\ref{sec:gwlensing} reviews wave-optics lensing and describes the evaluation of $F(w)$.

\subsection{Local lens model}
\label{sec:methods:fw}

We describe the lens in the thin-lens approximation and measure lens-plane positions $\bm x$ in units of a reference length $\xi_0$. The scaled source position, the impact parameter of the lens, is $\bm y = \bm\beta\, D_L/(\xi_0 D_S)$, with $\bm\beta$ the source position in the source plane and $D_L$, $D_S$, and $D_{LS}$ the angular diameter distances to the lens, to the source, and between the lens and the source. Hereafter we take
\begin{equation}
     \xi_0 = \ell_c\,,
\end{equation}
i.e.~the reference scale is set to the coherence length of the ULDM field. The arrival time of a ray crossing the lens plane at $\bm x$ is given by the dimensionless Fermat potential
\begin{equation}
\phi(\bm x,\bm y) = \tfrac12|\bm x-\bm y|^{2} - \psi(\bm x) + \phi_m(\bm y),
\label{eq:fermat}
\end{equation}
where $\psi$ is the dimensionless lensing potential and $\phi_m$ is chosen so that the minimum time delay vanishes (see Ref.~\cite{VillarrubiaRojo2024GLoW} for additional details). The lensing potential is sourced by the convergence through $\nabla_{\bm x}^{2}\psi=2\kappa$, where $\kappa=\Sigma/\Sigma_{\rm crit}$, $\Sigma_{\rm crit}=1/(4\pi G D_{\rm eff})$ is the critical surface density, and $D_{\rm eff}=D_LD_{LS}/D_S$.

The mean profile of the lens halo varies on a galactic scale, much larger than $r_F$ in the LISA band. The host therefore acts as a geometric optics lens and produces negligible diffraction of its own in this band. We describe the host near a macroimage by its local convergence $\kappa_0=\Sigma_{\rm sm}(\xi)/\Sigma_{\rm crit}$ and shear, which we treat as constant across the region used in the diffraction calculation. We measure the local lens plane position $\bm x$ from the center of the granule box in units of $\ell_c$. The constant deflection of the host is absorbed into the local scaled source position $\bm y$.

The granules enter as a fluctuating surface density $\delta\Sigma(\bm x)=\Sigma(\bm x)-\Sigma_{\rm sm}$ around this macroimage. They add a convergence $\delta\kappa(\bm x)=\delta\Sigma(\bm x)/\Sigma_{\rm crit}=\kappa_0\delta(\bm x)$ to the host convergence, where $\delta\equiv\delta\Sigma/\Sigma_{\rm sm}$ and $\Sigma_{\rm sm}$ is the locally constant mean surface density. We generate $\Sigma$ using the KL construction of Sec.~\ref{sec:uldm-stats:kl}, with the mean density profile along the line of sight to the unperturbed macroimage.

We generate the field on a square of physical side length $L_{\rm box}$, chosen so that the variation of the mean surface density from its value at the box center remains below the standard deviation of the granule surface density fluctuations throughout the square. We impose periodic boundary conditions, so the granule field repeats with period $L_{\rm box}$ along each axis. The corresponding dimensionless side length is $L=L_{\rm box}/\ell_c$.

In the local model, $\psi$ denotes the potential generated by $\delta\Sigma$, with $\nabla_{\bm x}^2\psi=2\delta\kappa$. We solve this equation in Fourier space on the periodic box. The host contribution is included through its convergence and shear in the quadratic Fermat potential $\phi_0$. The combined Fermat potential is
\begin{equation}
\begin{split}
\phi(\bm x)&=\phi_0(\bm x)-\psi(\bm x),\\
\phi_0(\bm x)&=\tfrac12\,\bm x^{\!\top}\!A\,\bm x-\bm y\cdot\bm x+\tfrac12|\bm y|^{2},\\
A&=\mathrm{diag}\,(1-\kappa_0+\gamma_0,\,1-\kappa_0-\gamma_0).
\end{split}
\label{eq:fermat-ext}
\end{equation}
Here $\gamma_0\equiv\sqrt{\gamma_1^2+\gamma_2^2}$ is the shear magnitude, and $A$ is written in diagonal form. The matrix $A$ specifies the local curvature of $\phi_0$, and the unperturbed macroimage lies at $\bm x^{*}=A^{-1}\bm y$. The additive constant in $\phi_0$ fixes an arbitrary origin of arrival time.

The eigenvalues of $A$ determine the macroimage type. Two positive eigenvalues give a Type~I minimum, opposite signs a Type~II saddle, and two negative eigenvalues a Type~III maximum. The smooth host magnification is
\begin{equation}
|\mu_0|=\frac{1}{|\det A|}, \qquad \det A=(1-\kappa_0)^{2}-\gamma_0^{2}.
\label{eq:detA}
\end{equation}
We use the combined Fermat potential in Eq.~\eqref{eq:fermat-ext} in the diffraction integral of Eq.~\eqref{eq:Fw-diffraction} to compute $F(w)$ for each granule realization, as described in Sec.~\ref{sec:methods:fourier}.

\subsection{SIS host and macroimage position}
\label{sec:methods:sis}

For the SIS profile in Eq.~\eqref{eq:sis-host-profile}, the local convergence is $\kappa_0=\xi_E/(2\xi)$, where $\xi_E=4\pi \sigma_v^2 D_{\rm eff}$ is the host Einstein radius in the lens plane. Axial symmetry allows us to choose the unperturbed macroimage at $\bm\xi_0=(\xi,0)$ in coordinates centered on the halo. With the local $x_1$ axis in the radial direction, $\gamma_0=\kappa_0=-\gamma_1$ and $\gamma_2=0$. Equations~\eqref{eq:fermat-ext} and~\eqref{eq:detA} then give
\begin{equation}
\begin{split}
A&=\mathrm{diag}\,(1,\,1-2\kappa_0),\\
\det A&=1-2\kappa_0, \qquad |\mu_0|=\frac{1}{|1-2\kappa_0|}.
\end{split}
\label{eq:mu0}
\end{equation}
The macroimage is a Type~I minimum for $\xi>\xi_E$ and a Type~II saddle for $\xi<\xi_E$. The critical curve lies at $\xi=\xi_E$, where $\det A=0$.

For the SIS, whose mean surface density decreases as $1/\xi$, the box size criterion above gives 
\begin{equation}
L_{\rm box}=\sqrt{\frac{\xi\ell_c}{\sqrt{\pi}}}\simeq\frac{\sqrt{2}\,\xi}{\sqrt{N_{\rm eff}}},
\label{eq:box-size-sis}
\end{equation}
where we used Eq.~\eqref{eq:neff-sis}. This gives the fiducial box sizes listed in Table~\ref{tab:physics-cases}. For these cases, $r_F<L_{\rm box}$ over $10^{-4}$--$10^{-1}$~Hz. Diffraction primarily samples the field within one period when the effective Fresnel region around the macroimage is well contained within the box. Section~\ref{sec:methods:fourier} discusses how periodic repetition affects the calculation when the effective Fresnel region approaches the box size.

The granule convergence has standard deviation $\sqrt{\operatorname{Var}(\delta\kappa)}=\kappa_0/\sqrt{N_{\rm eff}}$, with $N_{\rm eff}$ given by Eq.~\eqref{eq:neff-sis}. At fixed particle mass and lensing distances, this gives $\sqrt{\operatorname{Var}(\delta\kappa)}\propto\sigma_v^{3/2}\xi^{-3/2}$. Thus increasing $\sigma_v$ strengthens the convergence fluctuations at fixed $\xi$, while at fixed $\xi/\xi_E$ the scaling becomes $\sqrt{\operatorname{Var}(\delta\kappa)}\propto\sigma_v^{-3/2}$. The effect on $F(w)$ also depends on the host matrix $A$ and the frequency, as discussed in Sec.~\ref{sec:results:xi-scan}.

%%%%%
\section{Wave-Optics Gravitational Lensing}
\label{sec:gwlensing}
%%%%%
We first review the amplification factor and its regimes (Sec.~\ref{sec:gwlensing:Fw}), and then describe how we evaluate it for the lens of Sec.~\ref{sec:methods} (Sec.~\ref{sec:methods:fourier}).

\subsection{Amplification factor and diffraction regimes}
\label{sec:gwlensing:Fw}

A gravitational wave from a point-like, coherent source, lensed by a static thin lens, is described by a single complex, frequency-dependent amplification factor,
\begin{equation}
\tilde h(f) = F\big(w(f)\big)\,\tilde h_0(f), \label{eq:hlensed}
\end{equation}
where $\tilde h_0$ is the unlensed waveform and $F$ is given by the Fresnel--Kirchhoff diffraction integral over the lens plane (Eq.~\eqref{eq:Fw-diffraction} below)~\cite{DeguchiWatson1986,TakahashiNakamura2003,VillarrubiaRojo2024GLoW}.

The frequency enters only through the dimensionless combination
\begin{equation}
w(f) = \frac{2\pi f\,(1+z_L)\,\ell_c^{2}}{\,D_{\rm eff}}, \qquad D_{\rm eff} = \frac{D_L D_{LS}}{D_S},
\label{eq:w-of-f}
\end{equation}
where $z_L$ is the lens redshift and $D_{\rm eff}$ is the effective distance defined in Sec.~\ref{sec:methods:fw}. Introducing the Fresnel scale
\begin{equation}
r_F(f) = \sqrt{\frac{\,D_{\rm eff}}{2\pi f\,(1+z_L)}}, \label{eq:fresnel}
\end{equation}
Eq.~\eqref{eq:w-of-f} takes the transparent form
\begin{equation}
w = \left(\frac{\ell_c}{r_F}\right)^{2},
\label{eq:w-fresnel}
\end{equation}
so that $w$ compares the coherence scale of the lens with the Fresnel scale of the wave. Diffraction is maximal for $r_F\sim\ell_c$, that is, for $w\sim1$, and is suppressed for $w\ll1$, where the wave averages over the granules. The same criterion can be phrased in terms of mass. In terms of the critical surface density $\Sigma_{\rm crit}$, Eq.~\eqref{eq:w-of-f} becomes
\begin{equation}
w = 8\pi G M_{\rm eff}(1+z_L)\,f, \qquad M_{\rm eff}\equiv\pi\Sigma_{\rm crit}\,\ell_c^{2},
\label{eq:meff}
\end{equation}

The amplification factor is
\begin{equation}
F(w) = \frac{w}{2\pi i}\int d^{2}\bm x\; \exp\!\big[\,i\,w\,\phi(\bm x,\bm y)\,\big],
\label{eq:Fw-diffraction}
\end{equation}
where $\phi$ is the Fermat potential of Eq.~\eqref{eq:fermat}; for the lens of Sec.~\ref{sec:methods} it is given by Eq.~\eqref{eq:fermat-ext}.

The projected density is smooth on scales much smaller than $\ell_c$, owing to the exponential suppression of power in Eq.~\eqref{eq:projected-power}. Thus, when $r_F\ll\ell_c$, or equivalently $w\gg1$, the density varies little across the Fresnel scale and geometric optics applies, except near critical curves~\cite{EzquiagaLoVujeva2025}. In this regime the integral is dominated by the stationary points of $\phi$ and reduces to a sum over images weighted by their magnifications and time delays~\cite{TakahashiNakamura2003},
\begin{equation}
F(w)\simeq\sum_j\sqrt{|\mu_j|}\;e^{\,iwT_j}\,M_j,
\qquad
M_j=e^{-i\pi n_j/2}.
\label{eq:morse}
\end{equation}
Here $T_j=\phi(\bm x_j,\bm y)$ is the time delay of image $j$, $\mu_j=1/\det(\partial_a\partial_b\phi)$ its magnification, and $n_j=0$, $1$ and $2$ for an image at a minimum, saddle and maximum of $\phi$, which we call Type~I, II and III images. The Morse phase $M_j$ is the only trace of the image type in $F$: it is $1$ for a minimum and $e^{-i\pi/2}$ for a saddle. For a single image, Eq.~\eqref{eq:morse} is a constant amplitude and a phase linear in $w$, which are degenerate with the source distance, coalescence time and phase. Geometric optics is then unobservable, and only the departure from it carries information (Sec.~\ref{sec:results:setup}).

When $w\sim1$, the Fresnel scale is comparable to $\ell_c$, and diffraction by the granules produces frequency dependent modulations of the waveform amplitude and phase. For $w\ll1$, the much larger Fresnel region contains many granules, so their density fluctuations are averaged out. The granule induced modulations are consequently suppressed, and the amplification approaches that of the smooth lens halo; we make this statement precise in Sec.~\ref{sec:methods:modes}.

\subsection{Evaluation of the amplification factor}
\label{sec:methods:fourier}

Equation~\eqref{eq:Fw-diffraction} is numerically challenging, because its integrand oscillates increasingly rapidly as $w$ grows. The standard approach integrates along contours of equal Fermat potential, which turns it into a real, non-oscillatory function of the time delay whose one-dimensional Fourier transform gives $F(w)$ at all frequencies at once~\cite{Ulmer:1994ij,Diego:2019lcd,Tambalo2022WaveOptics,VillarrubiaRojo2024GLoW}. In the current public version of \texttt{GLoW}~\cite{VillarrubiaRojo2024GLoW}, the contour tracer starts from the global minimum and ends on an asymptotic family of contours at large $\phi$. A saddle macroimage has a different asymptotic contour topology when the external potential is constant. The available methods cannot be applied to Type~II images, and for Type~I images it fails to complete on about half of our realizations (Appendix~\ref{app:glow}).

\begin{figure*}[t]
\centering
\includegraphics[width=\textwidth]{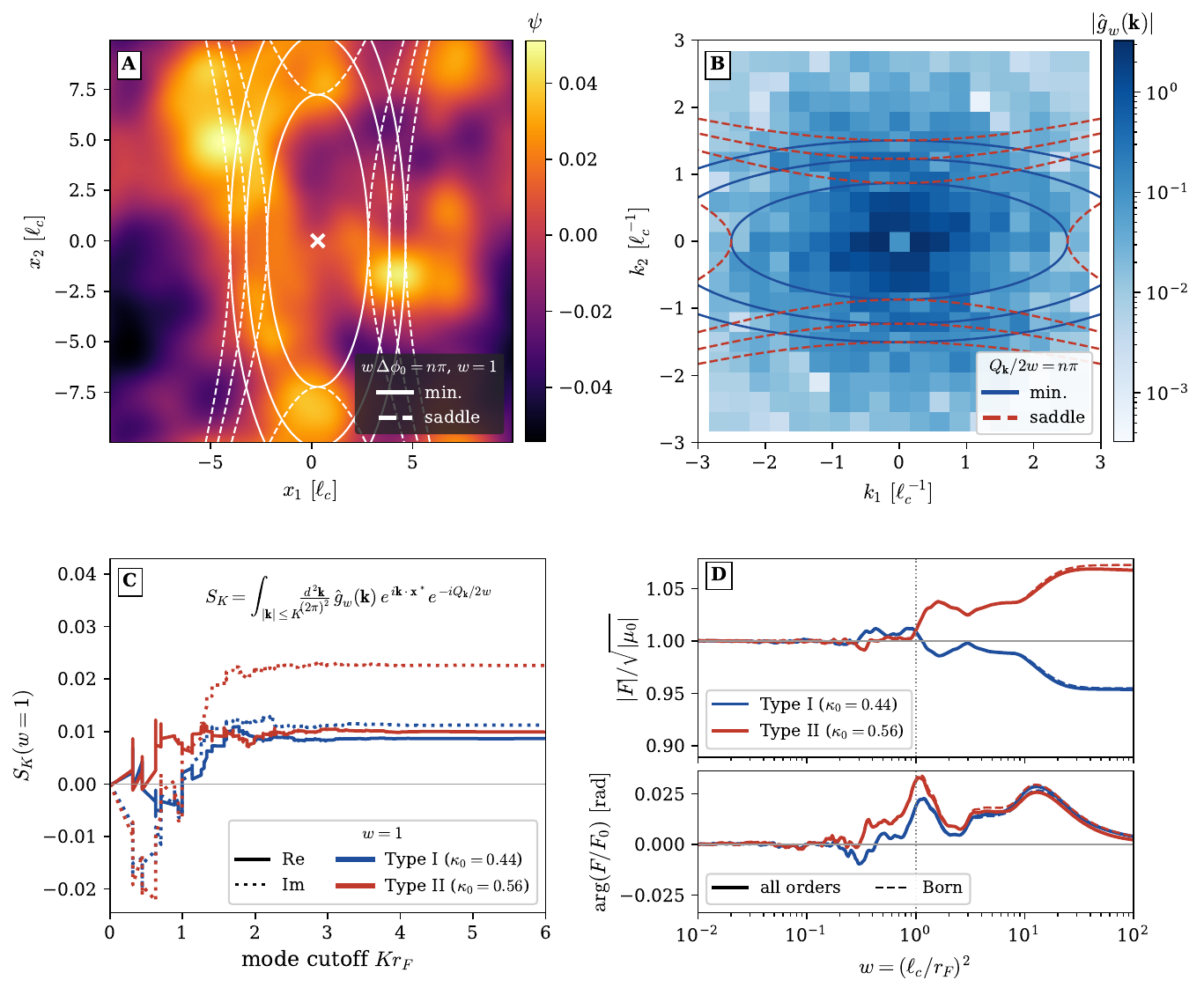}
\caption{Fourier-space evaluation of $F(w)$ for one granule realization (Halo~C, $m_\phi=10^{-21}$~eV), lensed as a Type~I minimum ($\kappa_0=0.44$) and a Type~II saddle ($\kappa_0=0.56$). Lengths and wave numbers are expressed in units of $\ell_c$ and $\ell_c^{-1}$, respectively. \emph{\textbf{A}}: granule potential $\psi$ over the periodic box, with the Fresnel zones $w\,\Delta\phi_0=n\pi$ of the macroimage (cross) at $w=1$, elliptic for the minimum (solid) and hyperbolic for the saddle (dashed). \emph{\textbf{B}}: spectrum $|\hat g_w(\bm k)|$ of the phase screen, Eq.~\eqref{eq:gw-fourier}, non-zero only on the lattice $\bm k\in(2\pi/L)\,\mathbb{Z}^2$ of the periodic box, with the transfer-function phases $Q_{\bm k}/2w=n\pi$ of Eq.~\eqref{eq:Qk}. \emph{\textbf{C}}: the $\bm k$ integral in Eq.~\eqref{eq:fourier-fresnel} restricted to the disk $|\bm k|\le K$, $S_K$, at $w=1$. It grows in steps because the lattice modes enter one shell at a time, and converges once $Kr_F\gtrsim2$--$3$. \emph{\textbf{D}}: amplitude and phase of $F/F_0$ across $w$, to all orders in $\psi$ (solid) and at Born order (dashed). For the phase, we remove the geometric image arrival time shift of the combined host and granule lens relative to the smooth host. At Born order, this removes the phase contribution $-w\psi(\bm x^*)$, where $\bm x^*$ is the unperturbed macroimage position.}
\label{fig:method}
\end{figure*}

We therefore evaluate Eq.~\eqref{eq:Fw-diffraction} by Fourier expanding the granule phase factor $e^{-iw\psi}$ over the lens plane. Our method is closely related to those presented in Refs.~\cite{Coles:2010da,Grillo:2018qjt,OguriTakahashi2022AmpPhase,Ephremidze:2026era}, and we discuss the relation once it has been derived (Sec.~\ref{sec:methods:modes}). The method rests on two properties of the composite lens in Eq.~\eqref{eq:fermat-ext}. The host part $\phi_0$ is quadratic, which gives a Gaussian integral that can be computed in closed form. Each plane wave adds only the linear term $\bm k\cdot\bm x$, preserving the quadratic form of the phase. This evaluation does not require locating the stationary points of the perturbed Fermat potential $\phi$, and applies to both minimum and saddle macroimages. We first evaluate the smooth host, and then add the granules mode by mode. Fig.~\ref{fig:method} illustrates the steps for one realization. The convergence of the mode sum and the comparison with \texttt{GLoW} are presented in Appendix~\ref{app:glow}.

After integration, each Fourier component contributes the smooth host amplification multiplied by a phase factor that depends on $\bm k$ and $w$. Here $\bm k$ is the spatial wavevector conjugate to the lens plane coordinate $\bm x$, while $w$ denotes the dimensionless GW frequency. We obtain $F(w)$ by summing these contributions with their complex Fourier coefficients, retaining the relative propagation phases between modes (Fig.~\ref{fig:method}A--B). In practice the sum converges once it includes all modes with $|\bm k|$ up to a few times $1/r_F$ (Fig.~\ref{fig:method}C). At higher frequencies, the phase factor $e^{-iw\psi}$ develops more rapid spatial variations, broadening its Fourier spectrum and requiring more modes (Appendix~\ref{app:glow}).

\subsubsection{The smooth host}
\label{sec:methods:F0}
With $\psi=0$, completing the square about the unperturbed image $\bm x^{*}\equiv A^{-1}\bm y$ gives
\begin{equation}
\begin{split}
\phi_0(\bm x)&=\tfrac12(\bm x-\bm x^{*})^{\!\top}\!A\,(\bm x-\bm x^{*})+t_0,\\
t_0&=\tfrac12|\bm y|^{2}-\tfrac12\,\bm y^{\!\top}\!A^{-1}\bm y ,
\end{split}
\label{eq:phi0-square}
\end{equation}
where $t_0$ is the time delay of the unperturbed image.

Since $A=\mathrm{diag}(a_1,a_2)$, with $a_1=1$ and $a_2=1-2\kappa_0$, the integral factorizes into one Fresnel integral along each axis,
\begin{equation}
\int_{-\infty}^{\infty}\!dv\;e^{\,iwav^{2}/2}=\sqrt{\frac{2\pi}{w|a|}}\;e^{\,i\,{\rm sgn}(a)\,\pi/4} .
\label{eq:fresnel-1d}
\end{equation}
Unlike a Gaussian integral with a real exponent, this oscillatory integral converges whatever the sign of $a$, as can be seen by inserting a regulator $e^{-\epsilon v^{2}}$ and letting $\epsilon\to0^{+}$; the sign enters only through the phase $\pm\pi/4$. It is this property that allows saddles to be treated exactly like minima.

Collecting the two factors together with the prefactor $w/(2\pi i)$ of Eq.~\eqref{eq:Fw-diffraction} gives
\begin{equation}
\begin{split}
F_0(w)&=\frac{1}{\sqrt{|a_1a_2|}}\;e^{\,i\frac{\pi}{4}[{\rm sgn}(a_1)+{\rm sgn}(a_2)]-i\frac{\pi}{2}}\;e^{iwt_0}\\
&=\sqrt{|\mu_0|}\;M\;e^{iwt_0} .
\end{split}
\label{eq:F0}
\end{equation}
Its modulus is the square root of the magnification of Eq.~\eqref{eq:mu0}, while its phase reduces to $-\pi n/2$, with $n$ the number of negative eigenvalues of $A$: the quadratic host reproduces the Morse phase $M$ of Eq.~\eqref{eq:morse}.

For $\kappa_0<1/2$ both eigenvalues of $A$ are positive, and the macroimage is a Type~I minimum with $n=0$ and $\mu_0=1/(1-2\kappa_0)$. For $\kappa_0>1/2$ the second eigenvalue changes sign, and the macroimage becomes a Type~II saddle with $n=1$, i.e.\ $M=e^{-i\pi/2}$, and $|\mu_0|=1/(2\kappa_0-1)$. Note that $|\mu_0|<1$ for $\kappa_0>1$, so a sufficiently strongly lensing host \emph{demagnifies} the saddle image and the baseline falls below unity.

For either parity, a constant $(\kappa,\gamma)$ field has a quadratic Fermat potential and is therefore not a localized lens. The smooth host thus contributes $\sqrt{|\mu_0|}$ at all frequencies, and the ULDM signature is a frequency-dependent modulation about this baseline.

\subsubsection{Adding the ULDM granules}
\label{sec:methods:modes}

Consider a single Fourier component of $e^{-iw\psi}$, represented by a plane wave $e^{i\bm k\cdot\bm x}$ multiplying the smooth host factor $e^{iw\phi_0}$ in Eq.~\eqref{eq:Fw-diffraction}.
Since its phase is linear in $\bm x$, it can be absorbed into the linear term of $\phi_0$,
\begin{equation}
w\,\phi_0(\bm x)+\bm k\cdot\bm x
=w\Big[\tfrac12\,\bm x^{\!\top}\!A\,\bm x-\Big(\bm y-\frac{\bm k}{w}\Big)\!\cdot\bm x+\tfrac12|\bm y|^{2}\Big] .
\end{equation}
which shows that each Fourier component of $e^{-iw\psi}$ shifts the source position in the Gaussian integral by $\bm y\to\bm y-\bm k/w$, leaving the curvature $A$ unchanged.

The smooth-host result of Sec.~\ref{sec:methods:F0} can therefore be reused with the same magnification and Morse phase; only the time delay has to be re-evaluated at the displaced source, the constant $\tfrac12|\bm y|^{2}$ being left as it is. This gives
\begin{equation}
\frac{w}{2\pi i}\int d^{2}\bm x\;e^{\,iw\phi_0(\bm x)}\,e^{i\bm k\cdot\bm x}=F_0(w)\;e^{\,i\bm k\cdot\bm x^{*}}\;e^{-iQ_{\bm k}/(2w)} ,
\label{eq:single-mode}
\end{equation}
with
\begin{equation}
Q_{\bm k}\equiv\bm k^{\!\top}\!A^{-1}\bm k=k_1^{2}+\frac{k_2^{2}}{1-2\kappa_0} .
\label{eq:Qk}
\end{equation}
The first phase, $\bm k\cdot\bm x^{*}$, is the value of the mode at the unperturbed image. The second, $-Q_{\bm k}/2w$ arises because the mode also moves the stationary point by $-A^{-1}\bm k/w$. It is the transfer function of the macroimage that appears in the Born-order kernel of Ref.~\cite{OguriTakahashi2022AmpPhase}, with their magnification eigenvalues $\mu_{j,i}=1/a_i$.

This second phase, $-Q_{\bm k}/(2w)$, describes diffraction through the relative phase shifts acquired by the plane wave components. 
Without the host, $A=\mathbf{1}$, it reduces to $Q_{\bm k}/(2w)=(kr_F/\ell_c)^2/2$, so it measures the spatial scale of the mode against the Fresnel scale of Eq.~\eqref{eq:fresnel}. Along the principal axes, the corresponding physical Fresnel scales are $r_F$ along $x_1$ and $r_F/|1-2\kappa_0|^{1/2}$ along $x_2$.
 
We write the granule factor of the integrand exactly as $e^{-iw\psi}=1+g_w$, subtracting the unperturbed term as in Eq.~(7) of Ref.~\cite{Ephremidze:2026era},
\begin{equation}
g_w(\bm x)\equiv e^{-iw\psi(\bm x)}-1 ,
\label{eq:gw}
\end{equation}
a function that can be written as a superposition of plane waves,
\begin{equation}
\begin{split}
g_w(\bm x)&=\int\!\frac{d^{2}\bm k}{(2\pi)^{2}}\;\hat g_w(\bm k)\,e^{i\bm k\cdot\bm x},\\
\hat g_w(\bm k)&=\int\!d^{2}\bm x\;g_w(\bm x)\,e^{-i\bm k\cdot\bm x}.
\end{split}
\label{eq:gw-fourier}
\end{equation}
Applying Eq.~\eqref{eq:single-mode} to each plane wave gives the central result of this section,
\begin{equation}
F(w)=F_0(w)\Big[\,1+\int\!\frac{d^{2}\bm k}{(2\pi)^{2}}\;\hat g_w(\bm k)\,e^{\,i\bm k\cdot\bm x^{*}}\,e^{-iQ_{\bm k}/(2w)}\Big].
\label{eq:fourier-fresnel}
\end{equation}
The integrand is the spectrum of the phase screen multiplied by the transfer function of the macroimage, $e^{-iQ_{\bm k}/2w}$, which dephases the modes with $Q_{\bm k}\gg2w$ (Fig.~\ref{fig:method}). On a map that is periodic with side $L$, $\hat g_w$ is non-zero only on the lattice $\bm k\in(2\pi/L)\,\mathbb{Z}^{2}$, and the integral becomes a sum, $\int d^{2}\bm k\,(2\pi)^{-2}\,\hat g_w\to L^{-2}\sum_{\bm k}\hat g_w(\bm k)$, which is evaluated by a single FFT of $g_w$. The periodic map also makes Eq.~\eqref{eq:fourier-fresnel} well defined: a homogeneous field extending over the whole plane does not decay at large $|\bm x|$, so its $\hat g_w$ exists only as a distribution, which on the periodic map reduces to this discrete set of modes.

Equation~\eqref{eq:fourier-fresnel} has transparent limits. At first order in $\psi$, $g_w\simeq-iw\psi$, and it reduces to the Born-order result of Ref.~\cite{OguriTakahashi2022AmpPhase} (Appendix~\ref{app:Fw:born}). The same expansion shows why the granules become invisible at low frequency: the correction is proportional to $w\psi$, the granule delay in units of the wave period. The transfer function has unit modulus and does not suppress it, but only redistributes it between amplitude and phase. Modes longer than the effective Fresnel scale, $Q_{\bm k}\ll2w$, contribute mostly to the phase, i.e.\ a time delay and a deflection degenerate with the source parameters, while the amplitude responds to the convergence of the granules averaged over the Fresnel zone, whose scatter scales as $\ell_c/r_F=w^{1/2}$, the inverse square root of the number of granules in the zone. At high frequency the phase of $g_w$ varies at the local rate $w|\nabla\psi|$, its spectrum extends to $|\bm k|\sim w|\nabla\psi|$, and these modes reconstruct the displacement of the image by the granule deflection, recovering geometric optics (Appendix~\ref{app:glow}). For the realization shown in Fig.~\ref{fig:method}D, the Born approximation closely reproduces the amplitude and phase modulations over the plotted frequency range. This agreement motivates using the Born approximation to interpret the ensemble statistics, which we examine in Appendix~\ref{app:Fw:born}.

We can now make explicit the relation to the works cited at the beginning of this section. Equation~\eqref{eq:fourier-fresnel} is the Fourier-space propagation of the phase screen $e^{-iw\psi}$ [Ref.~\cite{Grillo:2018qjt}, Eqs.~(A3)--(A6)], with the free-space Fresnel phase $k^{2}/2w$ replaced by that of the macroimage, $Q_{\bm k}/2w$; this transfer-function form has also been discussed for GW lensing in Ref.~\cite{Ephremidze:2026era} (Sec.~III\,A). The novelty of our approach resides in combining these elements: the transfer function of the macroimage, for minima and saddles alike, applied to all orders in $\psi$ on a periodic map.

Since no expansion in $\psi$ or in $1/w$ has been made, Eq.~\eqref{eq:fourier-fresnel} is exact for a periodic lensing potential, and evaluating it costs one FFT of $g_w$ per frequency on the grid of the $\psi$ map. On a $256^2$ grid this takes about $1$~ms per frequency on a single core of Apple M1, so $F(w)$ at our $750$ frequencies takes about $0.8$ s, and the result agrees with the contour integration of \texttt{GLoW} to $\lesssim10^{-3}$ wherever the latter applies (Appendix~\ref{app:glow}).

Generating the surface density maps costs more. Once the KL weights are computed~(Sec.~\ref{sec:uldm-stats:kl}), a map takes one inverse FFT per retained mode, $\mathcal{O}(nN^{2}\log N)$, against $\mathcal{O}(N_wN^{2}\log N)$ for $F(w)$ at $N_w$ frequencies, and for $n=3900$ on the same grid and core, one map takes $6.6$~s.

Two limitations remain. The first is resolution: how well that grid represents $g_w$, whose phase advances by $w|\nabla\psi|\Delta x$ from one cell to the next, which must stay well below $\pi$. Our grids have $\Delta x\simeq0.05\,\ell_c$, and the modes with $|\bm k|\le16/\ell_c$, well within the grid limit $\pi/\Delta x\simeq60/\ell_c$, already reproduce $F$ to $10^{-7}$ for $w\le100$ (Appendix~\ref{app:glow}). A lensing potential that varies over very small scales (e.g.\ a set of point lenses) will be inefficiently described by this method.
The second is periodicity: evaluated on a periodic map, Eq.~\eqref{eq:fourier-fresnel} describes an infinite lattice of identical granule boxes, the known constraint of phase-screen simulations~\cite{Coles:2010da}. At low $w$, where the effective Fresnel scale approaches the box size $L$, the periodic repetition can affect $F(w)$. We estimate the bias in the ensemble amplitude variance by comparing the Born predictions for the finite periodic box and the continuous power spectrum, as described in Appendix~\ref{app:Fw:born}.

%%%%%
\section{Detectability}
\label{sec:detectability}
%%%%%
We assess detectability at the single-event level, deferring any population-level questions to the conclusions (Sec.~\ref{sec:conclusions}).
We first specify the sources, the LISA sensitivity and the distinguishability criterion (Sec.~\ref{sec:results:setup}). We then show the amplification factor across ensembles of granule realizations for the benchmark hosts (Sec.~\ref{sec:results:amplification}), examine images close to the critical curve (Sec.~\ref{sec:results:near-critical}), and apply the criterion across the benchmark grid (Sec.~\ref{sec:results:detectability}).
Finally, we study how distinguishability varies with macroimage position on either side of the host critical curve (Sec.~\ref{sec:results:xi-scan}).

\subsection{Setup}
\label{sec:results:setup}

The lensed waveform is $\tilde h_{\rm L}(f)=F(w(f))\,\tilde h_{\rm U}(f)$ with $w(f)$ from Eq.~\eqref{eq:w-of-f}, at lens and source redshifts $z_L=1$ and $z_S=3$. Sources are equal-mass, non-spinning massive black hole binaries (MBHBs) of source-frame total mass $10^5$, $10^6$ and $10^7\,M_\odot$, modelled through inspiral, merger and ringdown with \textsc{IMRPhenomXAS}~\cite{Pratten2020IMRPhenomXAS}, which models only the dominant quadrupole harmonic, with $|\tilde h_{\rm U}(f)|^2$ averaged over inclination. Sensitivity is the sky- and polarization-averaged analytic LISA fit of Robson, Cornish and Liu~\cite{RobsonCornishLiu2019LISASensitivity} at a four-year mission duration. Each binary is integrated over the intersection of the LISA band with $[f(4\,{\rm yr\ before\ merger}),\,14f_{\rm ISCO}]$\footnote{Here $f_{\rm ISCO}$ is the GW frequency at the Schwarzschild innermost stable circular orbit for the redshifted total mass. The upper limit covers the full IMRPhenomXAS ringdown, which ends at $GMf=0.3$, approximately $13.85f_{\rm ISCO}$~\cite{Pratten2020IMRPhenomXAS}.}, giving unlensed optimal signal-to-noise ratios ${\rm SNR_{\rm U}}=379$, $2156$ and $577$ respectively. The lensed signal-to-noise ratio depends on the image, ${\rm SNR}_{\rm L}\simeq\sqrt{|\mu_0|}\,{\rm SNR}_{\rm U}$. For the saddles ($|\mu_0|=2.68$) and minima ($|\mu_0|=4.64$) of Fig.~\ref{fig:lindblom} it is $620$, $3530$ and $945$, and $817$, $4640$ and $1240$, respectively.

With the noise-weighted inner product $\langle a|b\rangle=4\,{\rm Re}\!\int\tilde a\tilde b^{*}/S_n(f)\,df$, we quantify the waveform distortion by the mismatch minimized over an arrival time shift $t_c$ and a constant phase $\varphi$,
\begin{equation}
\mathcal{M}=1-\max_{t_c,\varphi} \frac{\langle h_{\rm L}|h_{\rm U}\rangle} {\sqrt{\langle h_{\rm L}|h_{\rm L}\rangle\langle h_{\rm U}|h_{\rm U}\rangle}} .
\label{eq:mismatch}
\end{equation}

%Here $\varphi$ multiplies the positive frequency template by $e^{i\varphi}$. 
For the dominant quadrupole harmonic, optimizing over the constant phase is equivalent to varying the orbital phase. We define $\delta h=h_{\rm L}-h_{\rm U}^{\rm best}$, where $h_{\rm U}^{\rm best}$ is the reference template fitted in amplitude, arrival time and this constant phase. The two waveforms can be distinguished only if $\langle\delta h|\delta h\rangle\gtrsim1$, which for small mismatch gives $\mathcal{M}\,{\rm SNR}_{\rm L}^{2}\gtrsim1/2$~\cite{Lindblom2008WaveformAccuracy}. For our detectability analysis, we adopt the Lindblom criterion $\mathcal{M}\,{\rm SNR}_{\rm L}^{2}\gtrsim1$, following previous GW lensing studies~\cite{Savastano2023WOF,Brando2024WeaklyLensedGW}.

The frequency-dependent granule modulation produces the mismatch, while the host magnification changes the SNR at which this modulation is measured. The constant amplification $\sqrt{|\mu_0|}$ cancels from the normalized overlap, but gives ${\rm SNR}_{\rm L}^{2}=\langle h_{\rm L}|h_{\rm L}\rangle\simeq|\mu_0|\,{\rm SNR}_{\rm U}^{2}$, where ${\rm SNR}_{\rm U}^{2}=\langle h_{\rm U}|h_{\rm U}\rangle$. Magnification therefore helps resolve small granule modulations, while demagnification reduces their distinguishability. Our optimization also removes the constant Morse phase and an overall arrival time shift. 
%The host convergence and shear still affect the form of the granule modulation.

Figures~\ref{fig:lindblom} and~\ref{fig:xi-scan-xi-s} show $\mathcal{M}\,{\rm SNR}_{\rm L}^{2}$, while Fig.~\ref{fig:xi-scan} also shows $\mathcal{M}\,{\rm SNR}_{\rm U}^{2}$, which excludes the effect of host magnification on the SNR while retaining how the host convergence and shear modify the lensing response to the granules.

We hold the intrinsic source parameters fixed when comparing the lensed signal with the reference template. Allowing these parameters to vary could absorb part of the granule modulation and reduce the mismatch, so our detectability estimates are optimistic in this respect. Within this fixed parameter comparison, the source affects the mismatch through the distribution of its noise-weighted power, $\rho^{2}(f)=4|\tilde h_{\rm U}(f)|^{2}/S_n(f)$, across the integration band.

We tested the effect of higher harmonics on the distinguishability of the granule modulation by repeating the analysis with \textsc{IMRPhenomXHM}~\cite{GarciaQuiros2020IMRPhenomXHM}. We used the smooth-host waveform $F_0\,\tilde h_{\rm U}$ as template so that both signal and template include the host Morse phase. For each choice of source inclination and orbital phase, we maximized the overlap over the template's arrival time and orbital phase, keeping its inclination fixed to that of the source. We then averaged $\mathcal{M}\,{\rm SNR}_{\rm L}^{2}$ over the source inclinations and orbital phases. At every point of Fig.~\ref{fig:lindblom} the higher harmonics increase $\mathcal{M}\,{\rm SNR}_{\rm L}^{2}$: by less than $1\%$ for the $10^5\,M_\odot$ binary, by $4$--$29\%$ for the $10^6\,M_\odot$ binary, and by a factor of $1.3$--$2.2$ for the $10^7\,M_\odot$ binary. The threshold crossings move by at most $\sim10\%$ in $\sigma_v$, and the conclusions are unchanged.
\begin{figure*}[t]
\centering
\includegraphics[width=0.49\textwidth]{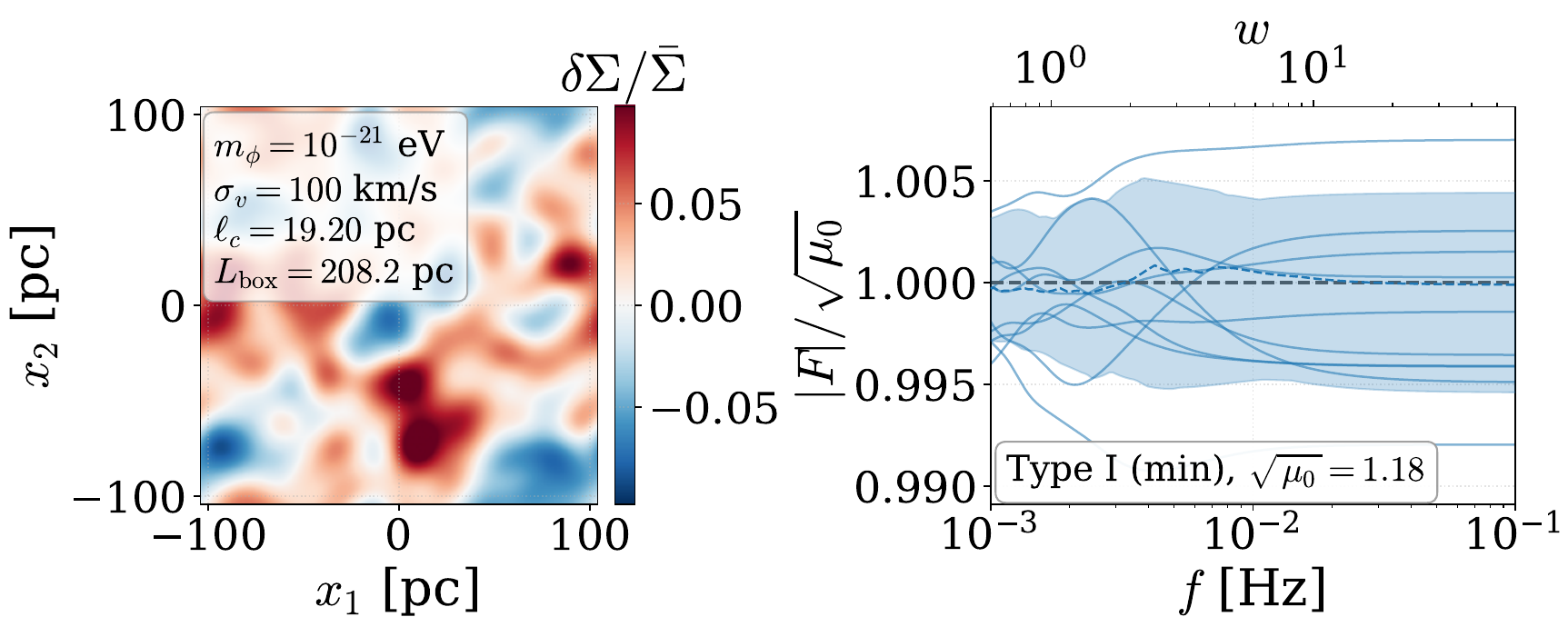}\hfill \includegraphics[width=0.49\textwidth]{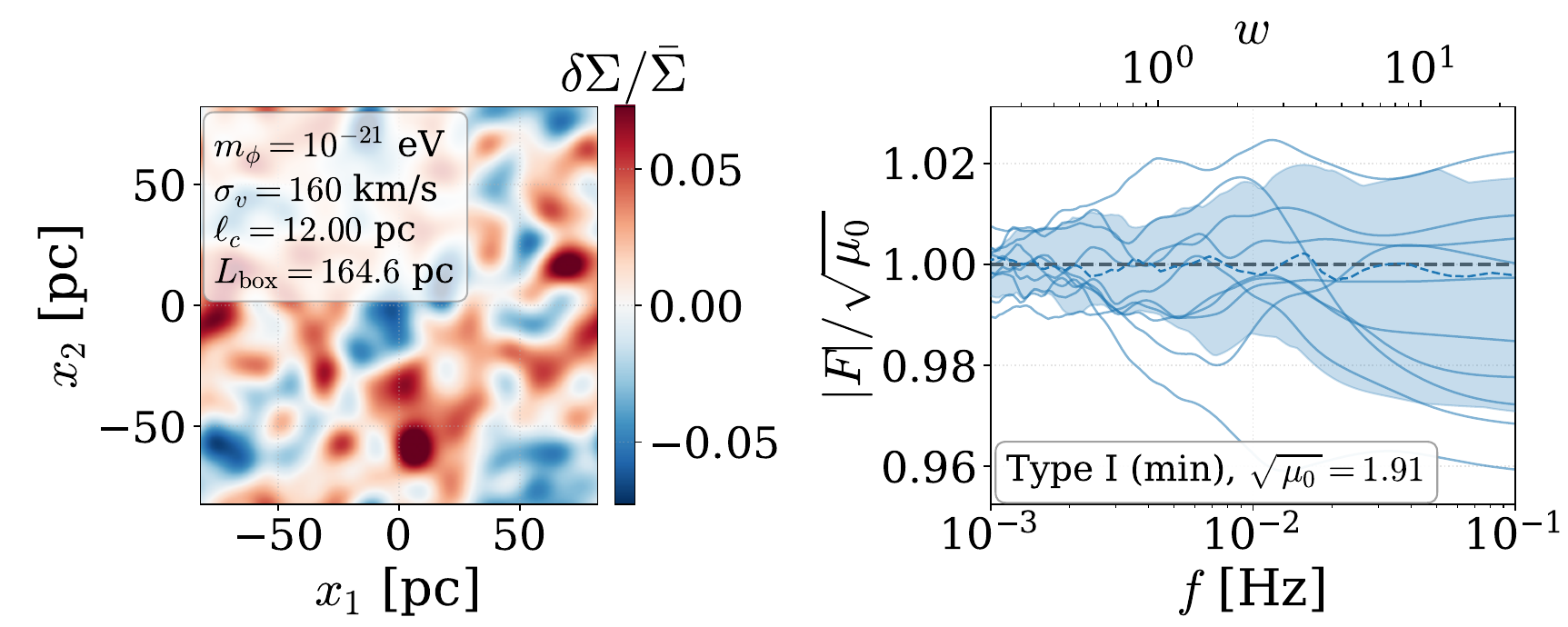}\\[2pt] \includegraphics[width=0.49\textwidth]{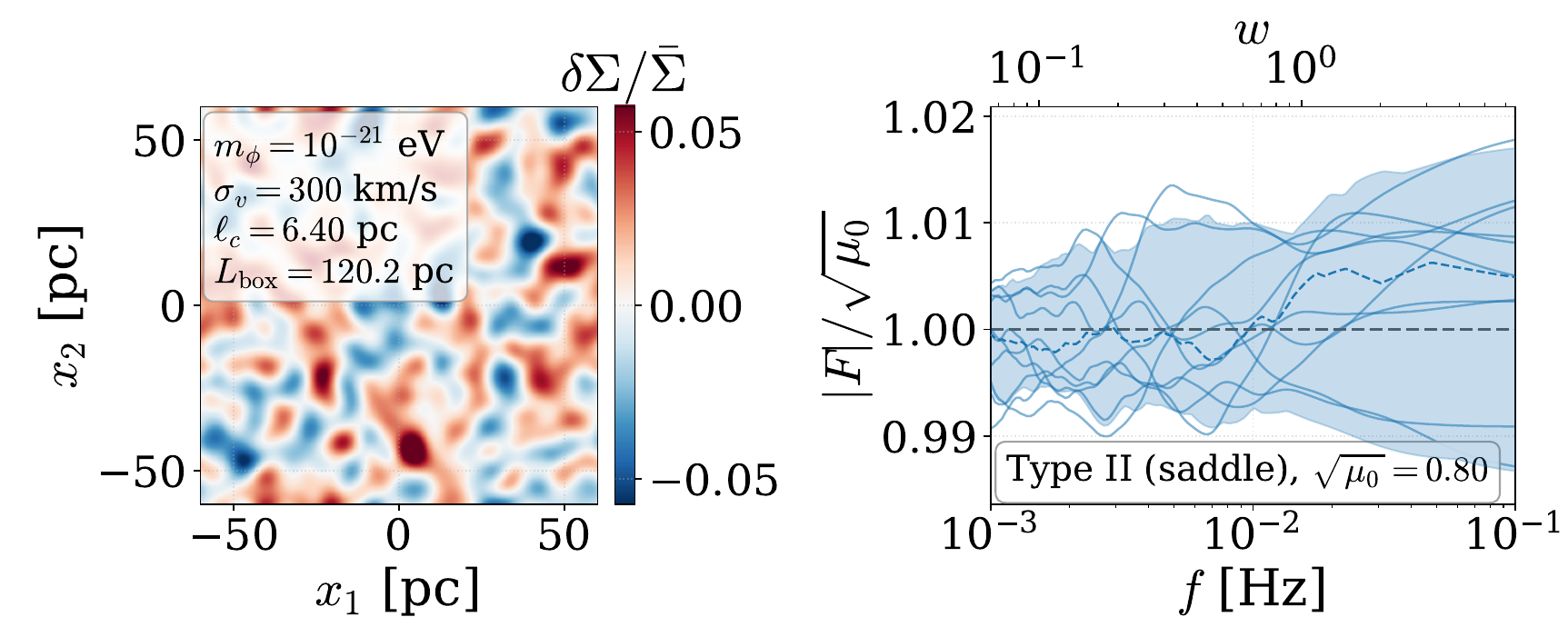}\hfill
\includegraphics[width=0.49\textwidth]{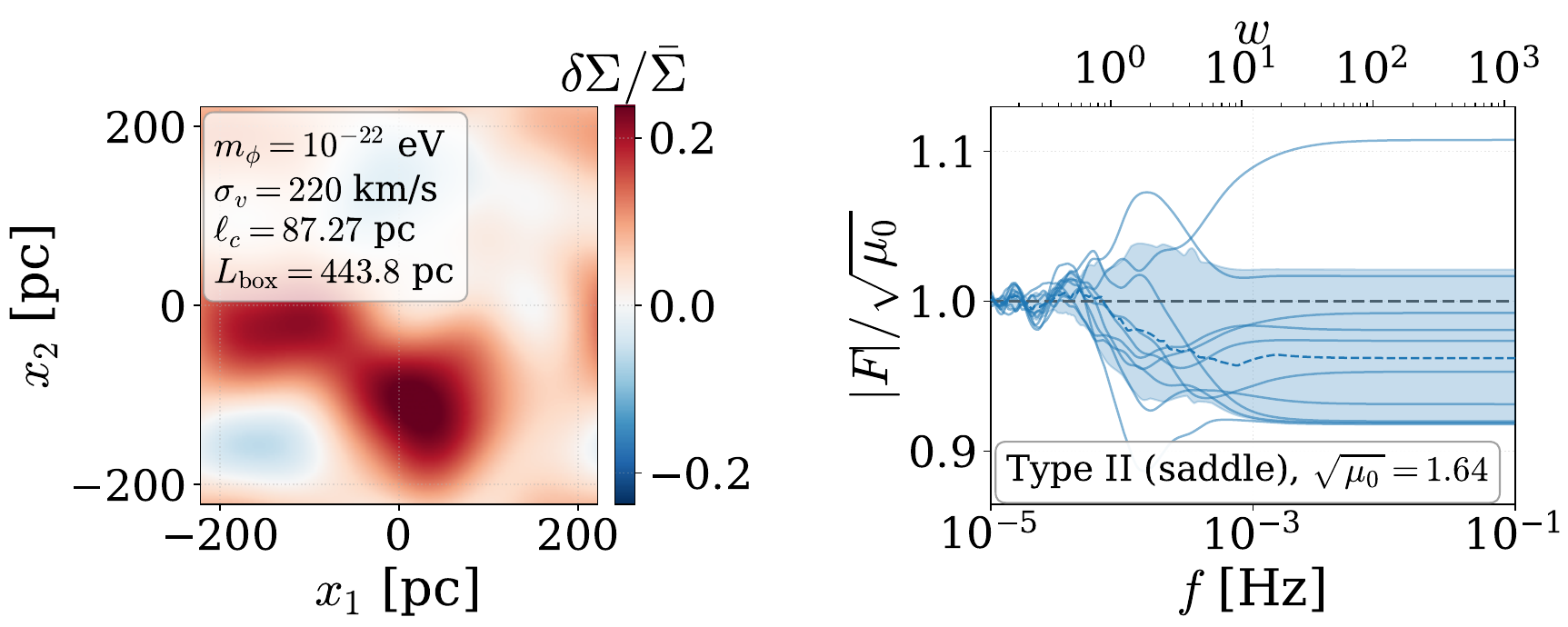}
\caption{Amplification $|F|$ due to ULDM granules, normalized by smooth host amplification $\sqrt{|\mu_0|}$, with baseline value quoted. Each panel shows one density-contrast map $\delta\Sigma/\Sigma_{\rm sm}$ (left) and $|F|/\sqrt{|\mu_0|}$ (right), where the blue dashed curve and shaded band give the median and $1\sigma$ range over $30$ realizations, thin lines show ten of these realizations, and the black dashed line the baseline. \emph{\textbf{Top.}} Type~I minima ($\sigma_v=100$ and $160$~km/s; $\kappa_0=0.142$ and $0.363$) at $m_\phi=10^{-21}$~eV. \emph{\textbf{Bottom.}} Type~II saddles: $\sigma_v=300$~km/s at $m_\phi=10^{-21}$~eV ($\kappa_0=1.276$) and $220$~km/s at $10^{-22}$~eV ($\kappa_0=0.686$). All panels are at $\xi=4$~kpc, where $\kappa_0\propto\sigma_v^{2}$ sets the parity; $m_\phi$, $\sigma_v$, $\sqrt{|\mu_0|}$, $\ell_c$ and $L_{\rm box}$ 
are quoted in each panel. The frequency corresponding to $w=1$ is lower for the larger $\ell_c$ at $10^{-22}$~eV ($f\approx8\times10^{-5}$~Hz).}
\label{fig:amplification}
\end{figure*}

\subsection{Lensing amplification} \label{sec:results:amplification}

Each panel of Fig.~\ref{fig:amplification} reports ensemble statistics of the granule lensing signal at fixed physical parameters. Across the simulated region, neighboring lines of sight share the same mean density profile, and the host convergence and shear are held constant. The granule field is statistically homogeneous in the lens plane, so the choice of source position $\bm y$ does not affect these statistics. We therefore compute $F(w)$ at the fixed source position $\bm y=(0.3,0)$ in units of $\ell_c$, in $30$ independently drawn field realizations. All panels use $z_L=1$, $z_S=3$, and the resulting effective distance $D_{\rm eff}=811.4$~Mpc.

Alongside the ensemble median and $1\sigma$ band, we overlay individual realizations to show how the modulation varies with frequency within a single realization. These curves retain the frequency-to-frequency correlations that the ensemble summaries at each frequency do not capture.

Two features are common to all four panels. At small $w$ the Fresnel scale $r_F$ far exceeds $\ell_c$, the granular structure is entirely unresolved, and only the macro-lens survives, so $|F(w)|$ sits on the smooth-host baseline $\sqrt{|\mu_0|}$ rather than on unity, a direct consequence of the external field being non-localized (Sec.~\ref{sec:methods:fourier}). Because each panel is normalized by its own baseline, the curves approach unity at low frequency.

Toward large $w$ the curves depart from the baseline and settle onto a geometric-optics plateau, offset from it by an amount that differs from one realization to the next, while the ensemble median stays within $1\%$ of the baseline at $m_\phi=10^{-21}$~eV and within $5\%$ at $10^{-22}$~eV. In between, at $w\sim0.1$--$3$, the granules imprint frequency-dependent modulations at the sub-percent to percent level, rising to $\approx6\%$ at $10^{-22}$~eV, where fewer patches along the line of sight leave a larger density contrast. This frequency dependence is the observable of interest, since the constant baseline cancels from the normalized overlap of Eq.~\eqref{eq:mismatch} (Sec.~\ref{sec:results:setup}).

The velocity-dispersion panels straddle the host's critical curve, with two Type~I minima and one Type~II saddle, but they differ in $\kappa_0$, $\mu_0$ and $\ell_c$ at the same time, so these panels alone do not separate the effect of parity from those of the host and the coherence length; Secs.~\ref{sec:results:near-critical} and \ref{sec:results:xi-scan} address this.

%%%%%
\subsection{Signal near the critical curve}
\label{sec:results:near-critical}
%%%%%
\begin{figure*}[tbp]
\centering \includegraphics[width=0.95\textwidth]{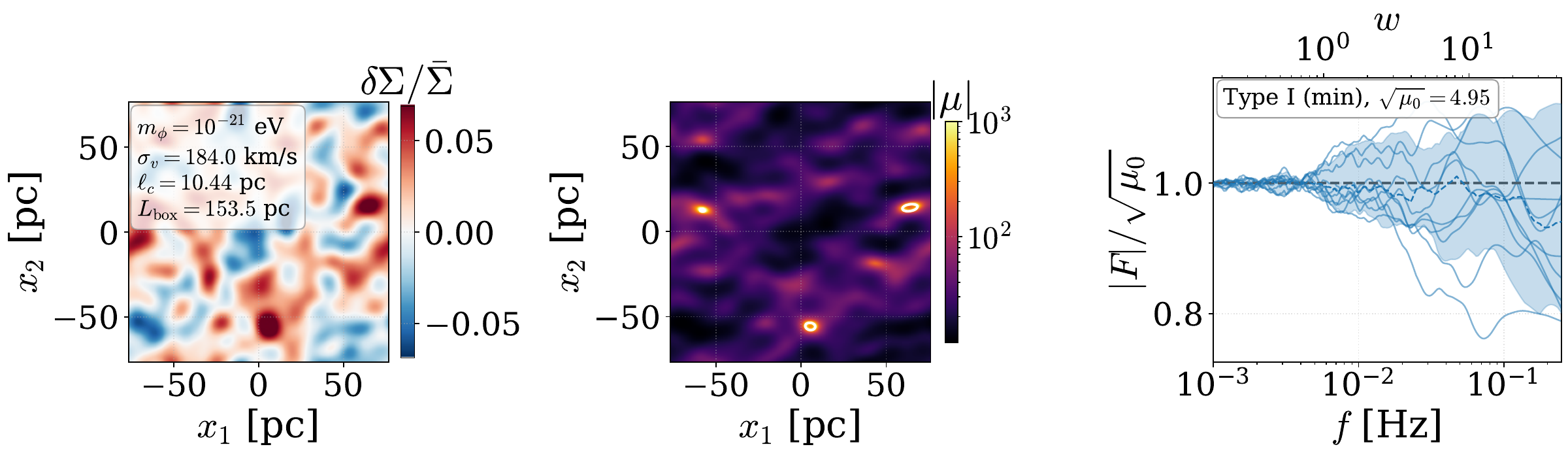}\\[2pt]
\includegraphics[width=0.95\textwidth]{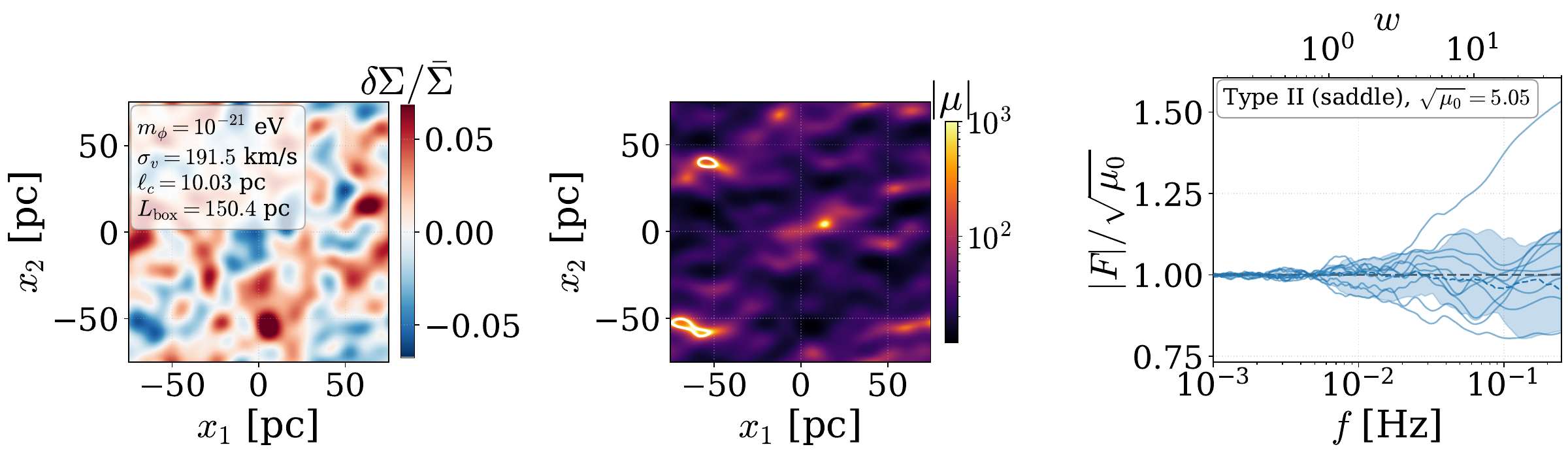}\\[2pt]
\includegraphics[width=0.95\textwidth]{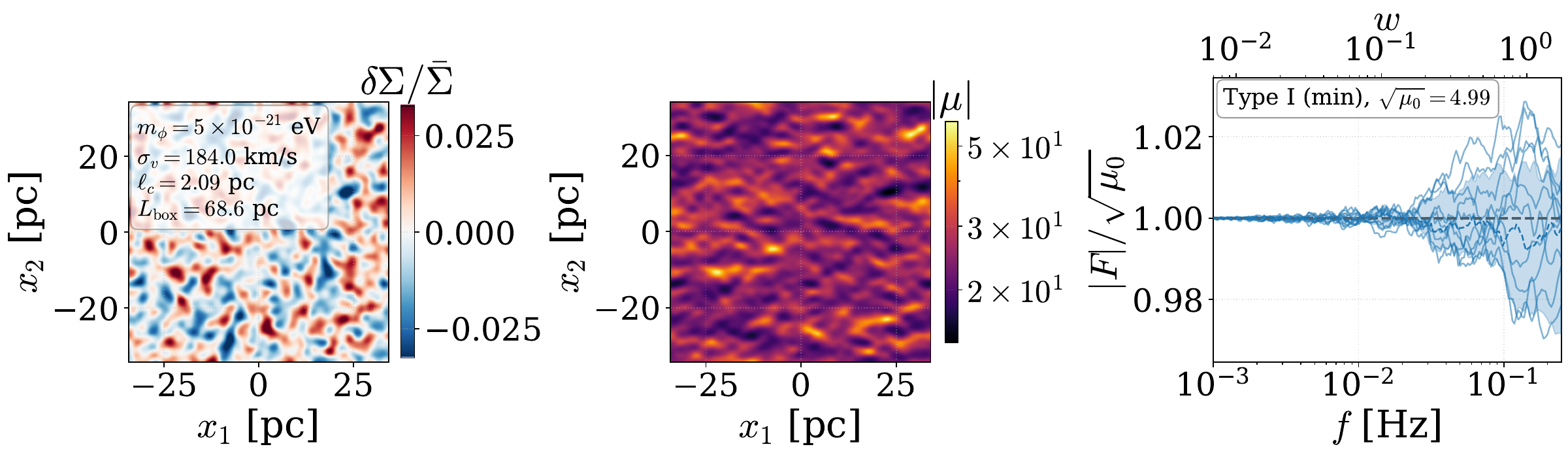}\\[2pt] \includegraphics[width=0.95\textwidth]{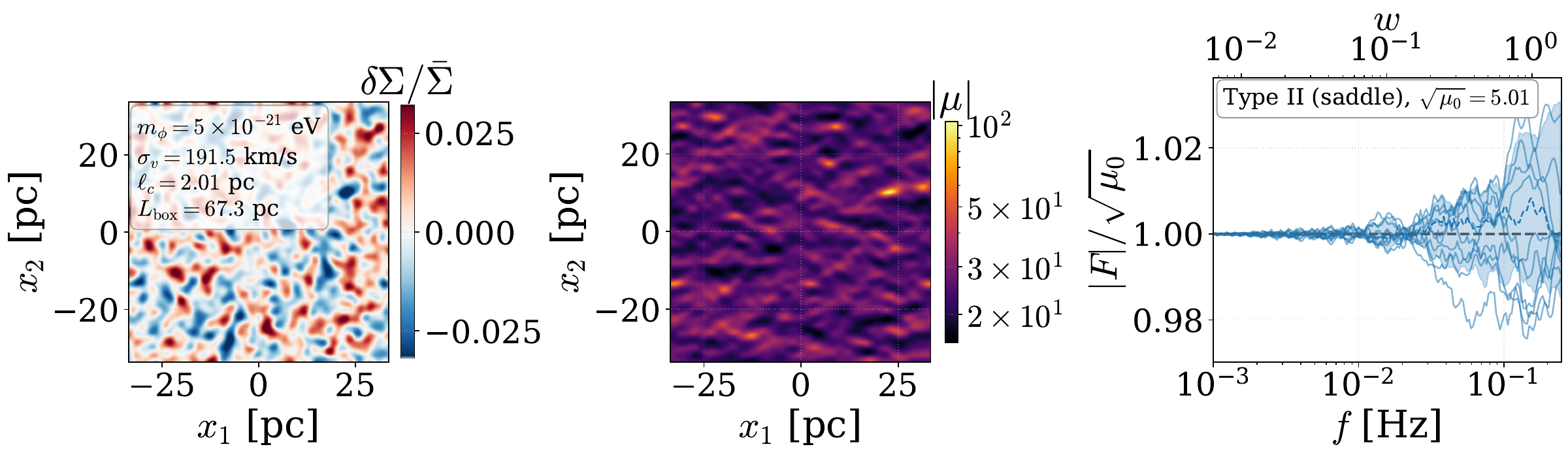}
\caption{Near-critical-curve amplification, both macroimage parities, for two ULDM masses at fixed $\xi=4$~kpc, $m_\phi\in\{1,5\}\times10^{-21}$~eV, at $z_L=1$, $z_S=3$. Rows are grouped by mass and alternate in parity. \emph{Rows~1 and 3} have $\sigma_v$ tuned to $\kappa_0=0.48$ (Type~I minimum, $\xi/\xi_E\approx1.042$, $\sigma_v=184.0$~km/s); \emph{rows~2 and 4} have $\sigma_v$ tuned to $\kappa_0=0.52$ (Type~II saddle, $\xi/\xi_E\approx0.962$, $\sigma_v=191.5$~km/s), the mirror image of the Type~I case across the critical curve. \emph{\textbf{Left.}} one density-contrast realization. \emph{\textbf{Middle.}} the geometric optics magnification map of the SIS host plus granules, using the same constant host convergence and shear as in the amplification calculation. The critical curves of the combined lens are overlaid in white. \emph{\textbf{Right.}} median and $1\sigma$ band of $|F|$ over $30$ realizations at a fixed probe point, with ten individual realizations shown as thin traces, each panel normalized to its own smooth-host baseline $\sqrt{|\mu_0|}$ so that the dashed baseline line lies at unity.
}
\label{fig:near-critical}
\end{figure*}

The granule signal is strongest for images close to the critical curve of the host. Figure~\ref{fig:near-critical} isolates this regime with a minimum and a saddle placed just outside and just inside it, each for $m_\phi=10^{-21}$ and $5\times10^{-21}$~eV. Both images sit at $\xi=4$~kpc, with $\sigma_v$ tuned to $\kappa_0=0.48$ and $0.52$, so that $|\mu_0|\simeq25$ in either case.

Consider the lighter mass first. At $m_\phi=10^{-21}$~eV the granule convergence, $\delta\kappa_{\rm rms}\approx0.013$, is comparable to the offset of $\kappa_0$ from its critical value $1/2$. The granules then do more than perturb the magnification: near the host critical curve, even small surface density fluctuations can create critical curves in otherwise noncritical regions (Fig.~\ref{fig:near-critical}: middle column, rows~1 and 2). A similar sensitivity to small-scale density structure underlies the microlensing of highly magnified stars near cluster caustics~\cite{Venumadhav2017Microcaustics,Dai:2020rio}.

At $m_\phi=10^{-21}$~eV, the ensemble standard deviation of $|F|/\sqrt{|\mu_0|}$ grows from $\approx2.5\%$ at $w=1$ to $10$--$12\%$ at $w=10$, several times the modulation of the benchmark hosts in Fig.~\ref{fig:amplification}.

The heavier mass, $5\times10^{-21}$~eV (Fig.~\ref{fig:near-critical}: rows~3 and 4), shrinks $\ell_c$ by a factor of five. The response as a function of $w$ is similar, but since $w\propto\ell_c^{2}f$, the diffractive window moves to frequencies $25$ times higher: $w=1$ falls at $f\approx0.15$~Hz rather than $6$~mHz. Over most of the LISA band, the modulation then stays below $\sim1\%$. The granule convergence, $\delta\kappa_{\rm rms}\approx0.006$, is also well below the offset of $\kappa_0$ from $1/2$, so the granules rarely create local critical curves. 

The enhanced granule signal requires an image near the critical curve, and hence a strongly lensed source. Depending on the populations of massive black hole binaries and lenses, LISA is expected to detect $\sim10^2$--$4\times10^4$ mergers in four years, of which $0.13$--$231$ are strongly lensed~\cite{Gutierrez:2025ymd}. For an SIS host, the magnification of both images near the critical curve is inversely proportional to the offset of the source from the lens axis, so the probability of a magnification above $|\mu|$ falls as $|\mu|^{-2}$~\cite{1992grle.book.....S}: about $1.5\%$ of strongly lensed sources produce an image with $|\mu_0|\gtrsim9$, as in Fig.~\ref{fig:overview-sis}, and only $\sim0.2\%$ one with $|\mu_0|\gtrsim25$, before magnification bias, which raises these fractions. Near-critical images are therefore the most favorable but rarest case.

%%%%%
\subsection{Distinguishability across the benchmark grid} \label{sec:results:detectability}
%%%%%
%%%% mass and sigma_v scan
\begin{figure*}[t]
\centering
\includegraphics[width=0.92\textwidth]{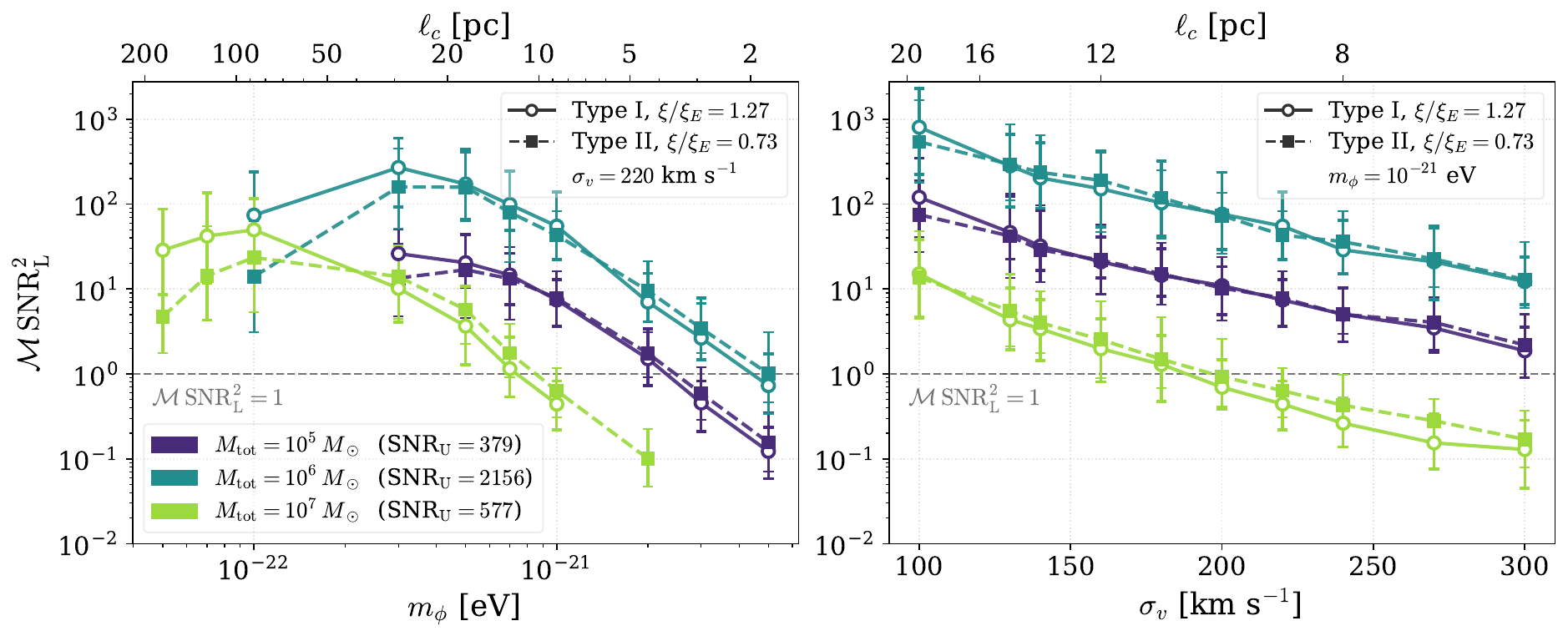}
\caption{Lindblom distinguishability $\mathcal{M}\,{\rm SNR}_{\rm L}^{2}$, with ${\rm SNR}_{\rm L}$ the signal-to-noise ratio of the lensed waveform, across the benchmark grid for all three MBHB sources, modelled with \textsc{IMRPhenomXAS} against the LISA noise curve of Ref.~\cite{RobsonCornishLiu2019LISASensitivity} over four years ($z_L=1$, $z_S=3$); the legend gives each benchmark's unlensed optimal ${\rm SNR}_{\rm U}$. Type~I minima are open circles with solid lines and Type~II saddles filled squares with dashed lines, in both panels. Saddles sit at $\xi/\xi_E=0.73$ ($\kappa_0=0.69$, $|\mu_0|=2.68$) and minima at $\xi/\xi_E=1.27$ ($\kappa_0=0.39$, $|\mu_0|=4.64$); \emph{\textbf{Left.}} ULDM mass scan at $\sigma_v=220$~km/s ($\xi=4$ and $7$~kpc). \emph{\textbf{Right.}} velocity-dispersion scan at $m_\phi=10^{-21}$~eV with $\xi=0.73\,\xi_E(\sigma_v)$ and $1.27\,\xi_E(\sigma_v)$. The secondary top axis gives the coherence length $\ell_c$ at each point. Points are medians and error bars the $1\sigma$ range over 60 realizations; the dashed line is the threshold $\mathcal{M}\,{\rm SNR}_{\rm L}^{2}=1$.}
\label{fig:lindblom}
\end{figure*}
Figure~\ref{fig:lindblom} shows $\mathcal{M}\,{\rm SNR}_{\rm L}^{2}$ across the benchmark grid. The threshold crossings discussed below refer to the ensemble medians. The mismatch itself is small throughout, $\mathcal{M}\approx3\times10^{-8}$--$2\times10^{-4}$, at the points shown, so the fitting-factor loss in recovered signal-to-noise is at most $\sim2\times10^{-4}$. Distinguishability is therefore carried by the large SNRs available to LISA rather than by a large waveform distortion.

The relative distinguishability of the granule modulation for the three sources depends on both their SNRs and how much of their noise-weighted power lies in the frequency range where $F$ varies. The $10^6\,M_\odot$ binary, whose band straddles $w\sim1$ at these coherence lengths, is the strongest probe over most of the grid. At the fiducial $m_\phi=10^{-21}$~eV, $\sigma_v=220$~km/s it reaches $\mathcal{M}\,{\rm SNR}_{\rm L}^{2}\approx43$ for the saddle and $55$ for the minimum, and it clears threshold at every velocity dispersion of the scan for both images, falling from $540$ (saddle) and $805$ (minimum) at $100$~km/s to $13$ and $12$ at $300$~km/s. Along the mass scan its saddle clears threshold from $m_\phi=10^{-22}$~eV to $5\times10^{-21}$~eV and its minimum up to $3\times10^{-21}$~eV, both peaking at $\approx160$ and $\approx270$, respectively, near $3\times10^{-22}$~eV. Since ${\rm SNR}_{\rm L}^{2}\simeq|\mu_0|\,{\rm SNR}_{\rm U}^{2}$, the minima, with $|\mu_0|=4.64$, gain more than the saddles, with $|\mu_0|=2.68$. Nevertheless, the two image types give comparable values of $\mathcal{M}\,{\rm SNR}_{\rm L}^{2}$ in the velocity dispersion scan, indicating that the saddles generally have larger mismatches at the positions considered here.

The $10^5\,M_\odot$ binary is quieter but comparably placed. Both its images clear the threshold for $3\times10^{-22}\le m_\phi\le2\times10^{-21}$~eV and at every velocity dispersion of the scan.

The $10^7\,M_\odot$ binary shows the role of band placement most clearly. Despite ${\rm SNR}_{\rm U}=577$, it fails for $m_\phi\ge10^{-21}$~eV at $\sigma_v=220$~km/s, because its lower merger frequency places most of its band at $w\lesssim0.1$, where the granules produce little frequency dependent modulation relative to the smooth host. The same placement favours it at the lightest masses, where $\ell_c$ is large: both its images clear the threshold for $m_\phi\le7\times10^{-22}$~eV. At $m_\phi\lesssim10^{-22}$~eV it is the strongest of the three sources for the saddle and comparable to the $10^6\,M_\odot$ binary for the minimum. In the velocity-dispersion scan, it clears the threshold up to $180$~km/s for both image types.

Realization-to-realization scatter is substantial: for a typical point the $1\sigma$ range of $\mathcal{M}\,{\rm SNR}_{\rm L}^{2}$ spans a factor of about six, so clearing the threshold describes some lines of sight rather than all.

\subsection{Dependence on macroimage position} \label{sec:results:xi-scan}

\begin{figure*}[t]
\centering \includegraphics[width=0.92\textwidth]{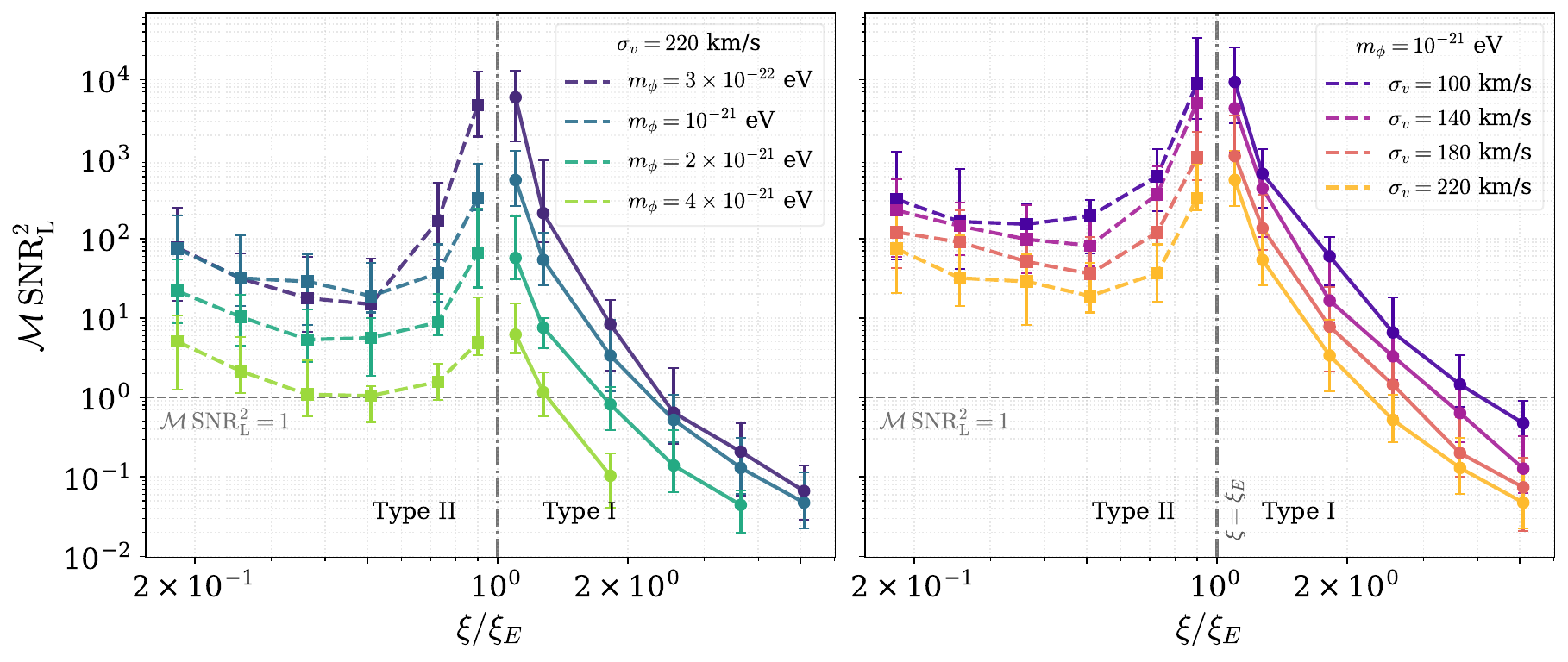} \caption{Lindblom distinguishability $\mathcal{M}\,{\rm SNR}_{\rm L}^{2}$ as a function of image position ($\xi/\xi_E$), including the host magnification. The GW source is assumed to be a $10^6\,M_\odot$ MBHB. \emph{\textbf{Left.}} ULDM mass scan, $m_\phi\in\{0.3,1,2,4\}\times10^{-21}$~eV at fixed $\sigma_v=220$~km/s. \emph{\textbf{Right.}} velocity-dispersion scan, $\sigma_v\in\{100,140,180,220\}$~km/s at fixed $m_\phi=10^{-21}$~eV. Points are medians with $1\sigma$ error bars over $30$ realizations; The two sides carry different macroimage parities: Type~II saddles ($\xi<\xi_E$, squares and dashed lines) and Type~I minima ($\xi>\xi_E$, circles and solid lines). The vertical dash-dotted line marks $\xi=\xi_E$ and the horizontal dashed line the Lindblom threshold $\mathcal{M}\,{\rm SNR}_{\rm L}^{2}=1$. Every member peaks close to the critical curve.}
\label{fig:xi-scan-xi-s}
\end{figure*}

\begin{figure*}[t] 
\centering \includegraphics[width=0.92\textwidth]{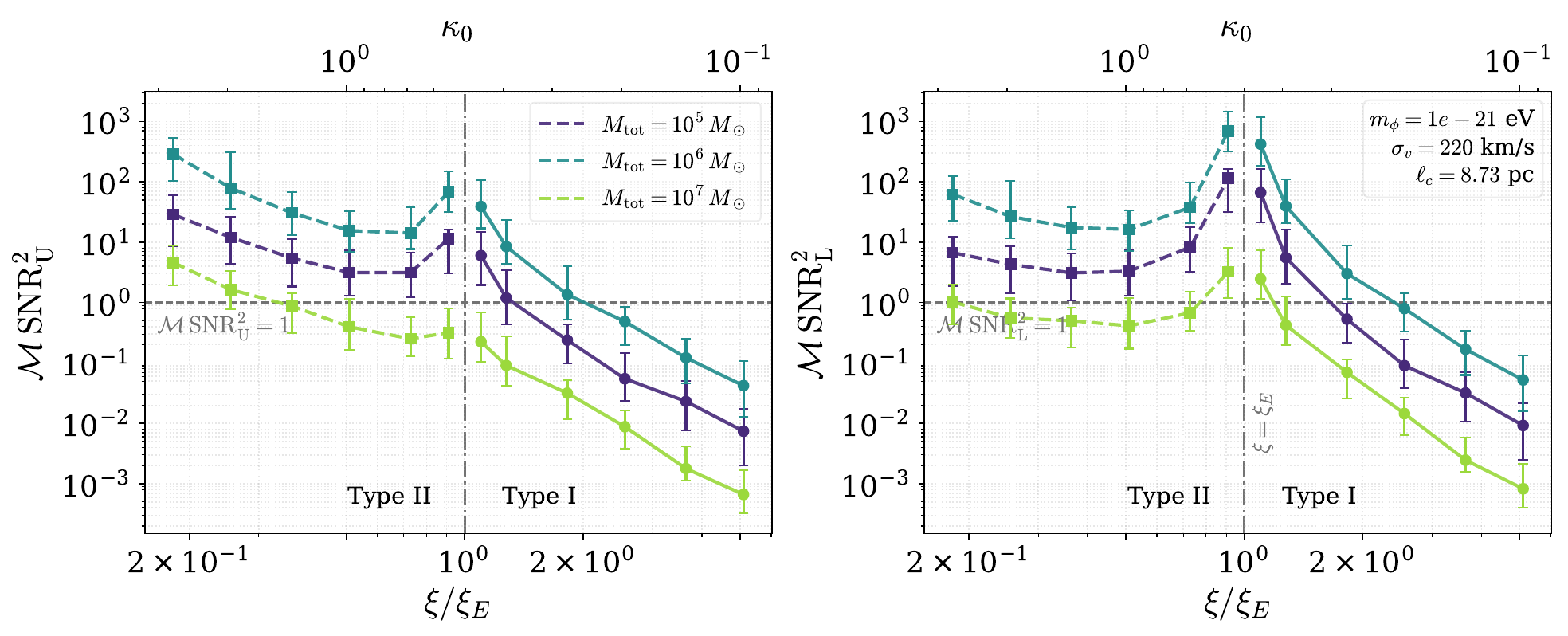} \caption{Distinguishability versus image position in units of the host Einstein radius, $\xi/\xi_E$, at fixed $m_\phi=10^{-21}$~eV and $\sigma_v=220$~km/s ($\ell_c=8.73$~pc), for all three MBHB benchmarks. 
\emph{\textbf{Left.}} $\mathcal{M}\,{\rm SNR}_{\rm U}^{2}$, using the unlensed SNR and excluding the effect of host magnification on the SNR. \emph{\textbf{Right.}} $\mathcal{M}\,{\rm SNR}_{\rm L}^{2}$, using the lensed SNR and including the host magnification. Both panels use the same mismatch, which includes how the host convergence and shear modify the lensing response to the granules. The vertical dash-dotted line marks $\xi=\xi_E$; the secondary top axis gives $\kappa_0=\xi_E/2\xi$. Type~II saddles ($\xi<\xi_E$) are shown with squares and dashed lines, Type~I minima ($\xi>\xi_E$) with circles and solid lines. Points show ensemble medians, and error bars indicate the $1\sigma$ range across granule realizations.}
\label{fig:xi-scan} 
\end{figure*}
Holding $(m_\phi,\sigma_v)$ fixed and sweeping $\xi$ moves the sightline toward or away from the host's Einstein radius. Figure~\ref{fig:xi-scan} scans $\xi=1$--$28$~kpc ($\xi/\xi_E=0.18$--$5.1$) for the benchmark host.

For each source, ${\rm SNR}_{\rm U}$ is independent of image position, so the left panel of Fig.~\ref{fig:xi-scan} directly shows the position dependence of the mismatch. The mismatch is non-monotonic for all three sources: it rises as $\xi$ decreases, peaks just inside $\xi_E$, dips at $\xi\simeq3$--$4$~kpc, then rises again toward the smallest $\xi$ probed. Including the host magnification (right panel of Fig.~\ref{fig:xi-scan}, $\mathcal{M}\,{\rm SNR}_{\rm L}^{2}$) multiplies each point by $\simeq|\mu_0|$. This sharpens the peak at the critical curve and lowers the points at $\xi/\xi_E<1/2$, where the saddles are demagnified.

This shape does not follow the granule amplitude, which grows monotonically inward (Sec.~\ref{sec:methods:sis}). It reflects how the host responds to the granules: close to the critical curve, where $1-2\kappa_0\to0$, the small curvature of the host Fermat potential along one direction enhances the response of $|F|$ to the granules (Appendix~\ref{app:Fw:born}), while the Fresnel phase $Q_{\bm k}/2w$ of Eq.~\eqref{eq:single-mode} changes the relative phases of the mode contributions differently on the two sides of the critical curve. The Born-order granularity signal of Fig.~\ref{fig:type-i-ii-summary} shows the same two features, a peak at the critical curve and a local minimum on the Type~II side.

Figure~\ref{fig:xi-scan-xi-s} extends this scan to two one-parameter families, one in $m_\phi$ and one in $\sigma_v$, in terms of $\mathcal{M}\,{\rm SNR}_{\rm L}^{2}$. The peak at the critical curve appears in every member of both, including the heaviest mass, $4\times10^{-21}$~eV: $\mathcal{M}\,{\rm SNR}_{\rm L}^{2}$ rises sharply from $\xi/\xi_E=0.73$ to the two points that bracket the transition, $0.900$ (Type~II) and $1.100$ (Type~I), and falls beyond them on either side, before rising again, more gently, toward small $\xi$ on the Type~II side.

%%%%%%%%%%%%%%%%%%%%%
\section{Conclusions}
\label{sec:conclusions}
%%%%%%%%%%%%%%%%%%%%%

We consider gravitational waves strongly lensed by a galaxy whose halo contains ultralight dark matter. As the waves travel through the halo, they encounter granules formed by interference of the ULDM field. These granules act as a diffractive barrier, forming a random screen of time delays whose spatial structure is set by the ULDM coherence length $\ell_c$. As the chirp sweeps the Fresnel scale across $\ell_c$, the statistics of the screen are imprinted into frequency dependent distortions of the waveform amplitude and phase. Previous studies exploring GW lensing signatures of ULDM have considered solitonic cores at halo centres as smooth lenses, while the effects of granules on the GW amplitude and phase have been studied only at Born order. We investigate the full wave-optics response to the granular field around strongly lensed minimum and saddle images, and whether its imprint can be distinguished in GW observations.

In this work we have computed this response for ULDM granules in a singular isothermal host. We approximated the surface density as a sum of independent contributions from coherent patches along the line of sight, with the patch depth set by the coherence length, to understand how projection changes the density statistics and when departures from a Gaussian distribution can matter. We generated surface density realizations from the covariance of the complex Gaussian ULDM field using a Karhunen--Lo\`eve expansion along the line of sight, allowing these departures to be retained in the lensing calculation (Sec.~\ref{sec:uldm-stats}). We described the host near each macroimage by its local convergence and shear, which covers minima and saddles alike (Sec.~\ref{sec:methods}), and evaluated the amplification factor to all orders in the granule potential by Fourier transforming the phase screen (Sec.~\ref{sec:gwlensing}). Finally, for ensembles of realizations spanning the boson mass, the velocity dispersion of the host and the image position, we computed the mismatch and the Lindblom distinguishability of massive black hole binaries observed by LISA (Sec.~\ref{sec:detectability}). Our main conclusions are:

\begin{itemize}
\item \emph{The ULDM signal is the frequency dependence of the amplification factor $F(f)$.} In our waveform model, a constant amplification and the Morse phase are absorbed into the source parameters. The granules modulate the amplitude and phase of the signal when $r_F\sim\ell_c$. At lower frequencies, the granule delay is a small fraction of the wave period, and the amplitude scatter decreases as $\ell_c/r_F$, as the Fresnel zone averages over more granules.

\item \emph{The frequency range of the modulations maps onto the boson mass.} Since $\ell_c=\hbar/(m_\phi\sigma_v)$, the modulation switches on at $f\simeq8\,{\rm mHz}\,(m_\phi/10^{-21}\,{\rm eV})^{2}(\sigma_v/220\,{\rm km\,s^{-1}})^{2}$ for $z_L=1$, $z_S=3$. For these lens parameters, the modulation falls within the LISA band for $m_\phi\sim10^{-22}$ to a few $\times10^{-21}$~eV. Increasing the boson mass by one or two orders of magnitude shifts the modulation into the decihertz or ground-based detector bands, respectively.

\item \emph{The surface density distribution of ULDM granules can depart from a Gaussian.} A longer coherence length or a shorter effective halo depth reduces the number of coherent patches contributing along the line of sight. Our coherent patch approximation quantifies the resulting departures from Gaussianity in terms of the mean halo profile and coherence length.

\item \emph{The macroimage position is crucial for the amplitude of the signal.} The modulation reaches the percent level for the benchmark hosts. Near the host critical curve, both minima and saddles become more sensitive to granule fluctuations, producing larger modulations (Sec.~\ref{sec:results:near-critical}). The granule fluctuations themselves also grow towards the halo centre. For an SIS host, saddles form deeper in the halo than minima of the same magnification and can therefore produce stronger modulations (Sec.~\ref{sec:results:xi-scan}), cf.\ the enhanced sensitivity of saddles to subhalos~\cite{Ando2026SaddleImages}.

\item \emph{LISA can distinguish the imprint.} Although the mismatch is small, the high signal-to-noise ratios of the strongly lensed massive black hole binaries expected in LISA~\cite{Sereno:2010dr,Gutierrez:2025ymd} make it distinguishable for suitable source and lens parameters at $m_\phi\sim10^{-21}$~eV.
\end{itemize}

More detailed forecasts can be achieved by going beyond some of our simplifying assumptions. The host is here described by its local convergence and shear, which is least accurate near the critical curve (Sec.~\ref{sec:results:near-critical}). The periodic box omits power on larger scales. The ULDM field has neither a solitonic core nor the correlations induced by self-gravity. Finally, the Lindblom criterion is only a necessary condition: parameter estimation must marginalize over the source, the smooth lens model and the granule realization, which may absorb part of the modulation.

The granule imprint could help distinguish ULDM from CDM. The percent level granule modulation in our strongly lensed benchmarks exceeds the sub-percent fluctuations expected from the line-of-sight population of $10$--$10^{4}\,M_\odot$ CDM halos in the LISA band~\cite{Choi2026LISAStochasticDiffraction}. In our model, the ULDM coherence length $\ell_c$ sets the characteristic granule scale and hence the frequency range of the modulation, whereas a CDM halo population contributes structure over a broad range of spatial scales. CDM subhalos near strongly lensed images also produce wave-optics imprints~\cite{OguriTakahashi2022AmpPhase,Ando:2026poq,Ando2026SaddleImages}, motivating a comparison within the same host to assess how well the signal strength and frequency dependence distinguish the two dark matter models.

The granule realization along a given sightline is unknown, so the imprint is best treated statistically. This is the idea behind stochastic diffraction, developed for the line-of-sight population of small halos and compact objects~\cite{Oguri:2020ldf,Zumalacarregui2024LSD,Urrutia:2024pos,Choi2026LISAStochasticDiffraction}, for the stars of a lens galaxy~\cite{Zumalacarregui:2026uqs}, and for ULDM granules around strongly lensed images~\cite{OguriTakahashi2022AmpPhase}: rather than modelling each lens, the data are compared with the predicted statistics of the frequency-dependent fluctuations of $F$. These statistics can be computed for any host, boson mass and image position, including granule maps from Schr\"odinger--Poisson simulations. Combined with realistic lens populations, this approach could turn strongly lensed GW events into a measurement of, or a bound on, $m_\phi$. These techniques may allow LISA and other upcoming GW detectors to act as diffraction microscopes to scrutinize the dark matter distribution on small scales.
%%%%%%%
\section*{Acknowledgments}
%%%%%%%
We are grateful to Ashish K. Meena, Sherry H. Suyu, Juan Urrutia, Claudi Vall M\"uller, Simona Vegetti, H\'ector Villarrubia-Rojo, and Simon D. M. White for discussions. MT thanks Nirmal Raj for providing access to computational resources, including NVIDIA A100 GPUs, which were used for some of the computations performed in this work.
MT acknowledges support from the Ministry of Education (MoE), India. Part of this work was carried out during two 2-month research visits to the Max Planck Institute for Gravitational Physics (Albert Einstein Institute, AEI) in Potsdam, Germany. MT thanks the Supporting members of the Max Planck Society for funding these visits through the grant ``Mit Einstein auf krummen Wegen'', and extends gratitude to the AEI Potsdam for its hospitality, assistance and scientific support during the research stay.
HGC acknowledges financial assistance from the Max Planck Society Supporting Members.
MZ acknowledges support from grant RYC2024-049805-I, funded by MICIU/AEI/10.13039/501100011033 and by the European Social Fund Plus (ESF+), and from the Ayudas de Excelencia RYC-MaX 2024 programme of the Spanish National Research Council (CSIC). This publication has been funded within the framework of the R\&D\&I Project CEX2025-001574-S, funded by MICIU/AEI/10.13039/501100011033. The research presented in this publication falls within the research line Origin and Composition of the Universe: Astroparticles and Cosmology (Astro/Cosmo). 
This work is supported by ERC grant GLOW (101230608 \url{www.glow-astro.org}) and  ERC grant ERC-2024-SYG 101167211. Funded by the European Union. Views and opinions expressed are however those of the author(s) only and do not necessarily reflect those of the European Union or the European Research Council Executive Agency. Neither the European Union nor the granting authority can be held responsible for them.
This publication is part of the R\&D\&i project PID2023-146686NB-C31 funded by MICIU/AEI/10.13039/501100011033/ and by ERDF/EU.
IFAE is partially funded by the CERCA program of the Generalitat de Catalunya.
D.B. acknowledges financial support from the Spanish Ministry of Science and Innovation (MICINN) through the Spanish State Research Agency, under Severo Ochoa Centres of Excellence Programme 2025-2029 (CEX2024001442-S).

\appendix
\renewcommand{\thesubsection}{\thesection.\arabic{subsection}}
\makeatletter
\renewcommand{\p@subsection}{}
\makeatother

\section{ULDM field statistics and surface density construction}
\label{app:uldm-statistics}

\subsection{Field statistics and covariance}
\label{app:uldm-wave}

With the envelope normalization in Eq.~\eqref{eq:uldm-envelope}, the nonrelativistic field equations are~\cite{HuBarkanaGruzinov2000,HuiOstrikerTremaineWitten2017}
\begin{equation}
i\partial_t\Psi=-\frac{\nabla^2\Psi}{2k_\phi}+k_\phi\Phi\Psi,
\qquad
\nabla^2\Phi=4\pi G\rho.
\label{eq:schrodinger-poisson}
\end{equation}
A constant potential contributes a common phase, and a uniform acceleration can be absorbed into a freely falling frame. The remaining tidal evolution is small over a coherence time when $\|\nabla\nabla\Phi\|\tau_c^2\ll1$. For density fluctuations of order $\rho_{\rm sm}$, the estimate $4\pi G\rho_{\rm sm}\tau_c^2\ll1$ ensures a slow gravitational response on that timescale. These conditions specify the local free wave approximation~\cite{Bar-Or:2018pxz}.

In a locally homogeneous normalization volume $V$, write $\Psi(t,\bm x)=V^{-1/2}\sum_{\bm k}\psi_{\bm k}e^{i\bm k\cdot\bm x-ik^2t/(2k_\phi)}$. We use independent complex Gaussian coefficients with distribution~\cite{Centers:2019dyn,Hui:2020hbq}
\begin{equation}
p(\{\psi_{\bm k}\})=
\prod_{\bm k}\frac{1}{\pi S_\Psi(\bm k)}
\exp\left[-\frac{|\psi_{\bm k}|^2}{S_\Psi(\bm k)}\right].
\label{eq:uldm-mode-pdf}
\end{equation}
Here $S_\Psi(\bm k)=\langle|\psi_{\bm k}|^2\rangle$, so $\langle|\Psi|^2\rangle=V^{-1}\sum_{\bm k}S_\Psi(\bm k)\to\int d^3k\,S_\Psi(\bm k)/(2\pi)^3$. Using $\bm v=\bm k/k_\phi$ and $\langle\rho\rangle=\rho_0$ gives
\begin{equation}
S_\Psi(\bm k)=\frac{(2\pi)^3\rho_0}{k_\phi^5}
f\left(\frac{\bm k}{k_\phi}\right),
\qquad
\int d^3v\,f(\bm v)=1.
\label{eq:wave-spectrum-velocity}
\end{equation}
At one point, the joint density is $p(\Re\Psi,\Im\Psi)=k_\phi^2(\pi\rho_0)^{-1}\exp[-k_\phi^2|\Psi|^2/\rho_0]$. Integrating over the phase and changing variables to $\rho=k_\phi^2|\Psi|^2$ gives Eq.~\eqref{eq:rho-exponential}.

The freely evolving Fourier modes give the field covariance
\begin{equation}
C_\Psi(\tau,\bm r)=\frac{\rho_0}{k_\phi^2}
\int d^3v\,f(\bm v)
e^{ik_\phi\bm v\cdot\bm r-ik_\phi v^2\tau/2},
\label{eq:psi-velocity-cov}
\end{equation}
where $\tau=t-t'$ and $\bm r=\bm x-\bm x'$. For the isotropic Maxwellian distribution
\begin{equation}
f(\bm v)=\frac{e^{-v^2/(2\sigma_v^2)}}{(2\pi\sigma_v^2)^{3/2}},
\label{eq:uldm-maxwell}
\end{equation}
the Gaussian integral gives
\begin{equation}
C_\Psi(\tau,r)=\frac{\rho_0}{k_\phi^2}
\left(1+i\frac{\tau}{\tau_c}\right)^{-3/2}
\exp\left[-\frac{r^2}{2\ell_c^2(1+i\tau/\tau_c)}\right].
\label{eq:maxwell-amplitude-cov}
\end{equation}
Wick contraction gives $\langle\rho\rho'\rangle=\rho_0^2+k_\phi^4|C_\Psi|^2$, yielding Eq.~\eqref{eq:drho-2pt}~\cite{Bar-Or:2018pxz,Hui:2020hbq}. Its spatial Fourier transform, defined by $C_{\delta\rho}(\tau,\bm r)=\int d^3k\,P_{\delta\rho}(k;\tau)e^{i\bm k\cdot\bm r}/(2\pi)^3$, is
\begin{equation}
P_{\delta\rho}(k;\tau)=\pi^{3/2}\rho_0^2\ell_c^3
\exp\left[-\frac{k^2\ell_c^2}{4}
\left(1+\frac{\tau^2}{\tau_c^2}\right)\right].
\label{eq:uldm-density-power}
\end{equation}

\subsection{Projected covariance and coherent patch statistics}
\label{app:uldm-projection}

Wick's theorem gives $C_{\delta\Sigma}=k_\phi^4\int dz\,dz'\,|C_\Psi|^2$. Substituting Eq.~\eqref{eq:psi-inhomogeneous} yields
\begin{equation}
\begin{split}
C_{\delta\Sigma}(\bm\xi,\bm\xi')={}&
e^{-|\bm\xi-\bm\xi'|^2/\ell_c^2}\\
&\times\int dz\,dz'\,
\rho_{\rm sm}(\bm\xi,z)\rho_{\rm sm}(\bm\xi',z')
e^{-(z-z')^2/\ell_c^2}.
\end{split}
\label{eq:surface-covariance-exact}
\end{equation}
For a mean profile varying slowly over $\ell_c$, evaluating the smooth factors at $z'=z$ and integrating over $z'-z$ gives Eq.~\eqref{eq:surface-variance-local} at coincident transverse positions. When neighboring lines of sight have nearly identical mean density profiles, the transverse Fourier transform gives Eq.~\eqref{eq:projected-power}.

The normalized longitudinal density covariance defines the effective patch depth,
\begin{equation}
\Delta z_{\rm eff}\equiv
\int_{-\infty}^{\infty}du\,e^{-u^2/\ell_c^2}
=\sqrt{\pi}\ell_c.
\label{eq:patch-depth}
\end{equation}
At a fixed $\bm\xi$, represent the surface density by independent exponential patch contributions,
\begin{equation}
\begin{split}
\Sigma&\simeq\sum_j a_jX_j,
\qquad
a_j=\rho_{\rm sm}(\bm\xi,z_j)\Delta z_{\rm eff},\\
p(X_j)&=e^{-X_j},\qquad X_j\geq0.
\end{split}
\label{eq:patch-sum}
\end{equation}
Here $\langle\cdot\rangle_{\rm patch}$ and $\operatorname{Var}_{\rm patch}$ denote ensemble averages and variances over the independent patch variables $X_j$, with the mean density profile and $\Delta z_{\rm eff}$ held fixed. The unit mean and unit variance of $X_j$ give
\begin{equation}
\begin{split}
\langle\Sigma\rangle_{\rm patch}
&=\sum_j a_j\simeq\int dz\,\rho_{\rm sm}=\Sigma_{\rm sm},\\
\operatorname{Var}_{\rm patch}\Sigma
&=\sum_j a_j^2\simeq\Delta z_{\rm eff}\int dz\,\rho_{\rm sm}^2.
\end{split}
\label{eq:patch-mean-variance}
\end{equation}
Thus Eq.~\eqref{eq:patch-depth} reproduces the leading variance in Eq.~\eqref{eq:surface-variance-local}, and Eq.~\eqref{eq:neff} becomes $N_{\rm eff}\simeq(\sum_j a_j)^2/\sum_j a_j^2$.

The probability density of the patch sum can be written using the Dirac delta $\delta_{\rm D}$. Substituting its Fourier representation and using the independence of the $X_j$ gives
\begin{equation}
\begin{split}
p_{\rm patch}(\Sigma|\bm\xi)
&=\left\langle\delta_{\rm D}\left(\Sigma-\sum_j a_jX_j\right)\right\rangle_{\rm patch}\\
&=\int_{-\infty}^{\infty}\frac{ds}{2\pi}\,e^{is\Sigma}
\prod_j\left\langle e^{-is a_jX_j}\right\rangle_{\rm patch}.
\end{split}
\label{eq:patch-pdf-fourier}
\end{equation}
For the exponential distribution in Eq.~\eqref{eq:patch-sum}, each factor is $\langle e^{-is a_jX_j}\rangle_{\rm patch}=\int_0^\infty dX_j\,e^{-(1+is a_j)X_j}=(1+is a_j)^{-1}$. Since $\Sigma=\sum_j a_jX_j$ in the discrete patch model, independence gives
\[
\left\langle e^{-is\Sigma}\right\rangle_{\rm patch}
=\prod_j\left\langle e^{-is a_jX_j}\right\rangle_{\rm patch}
=\prod_j(1+is a_j)^{-1}.
\]
Taking the logarithm turns this product into a sum over patches. Each patch has depth $\Delta z_{\rm eff}$ and mean surface density contribution $a_j=\Delta z_{\rm eff}\rho_{\rm sm}(\bm\xi,z_j)$, where $z_j$ is its centre. When $\rho_{\rm sm}$ varies little across a patch,
\[
\Delta z_{\rm eff}\ln(1+is a_j)
\simeq\int_{\text{patch }j}dz\,
\ln\left[1+is\Delta z_{\rm eff}\rho_{\rm sm}(\bm\xi,z)\right].
\]
Summing over the patches therefore converts the discrete sum into a line-of-sight integral,
\begin{equation}
\begin{split}
\ln\left\langle e^{-is\Sigma}\right\rangle_{\rm patch}
&=-\sum_j\ln(1+is a_j)\\
&\simeq-\frac{1}{\Delta z_{\rm eff}}
\int dz\,\ln\left[1+is\Delta z_{\rm eff}\rho_{\rm sm}(\bm\xi,z)\right].
\end{split}
\label{eq:patch-characteristic}
\end{equation}
Fourier inversion gives Eq.~\eqref{eq:patch-pdf-general}, with the logarithm continued from its value at $s=0$. The moment expressions below assume finite integrals $\int dz\,\rho_{\rm sm}^p$ for $p=1,\ldots,4$.

To control terms beyond quadratic order in Eq.~\eqref{eq:patch-log-quadratic}, expand the Fourier exponent,
\begin{equation}
\begin{split}
&is\Sigma-\frac{1}{\Delta z_{\rm eff}}\int dz\,\ln(1+is\Delta z_{\rm eff}\rho_{\rm sm})\\
&\quad\simeq is(\Sigma-\Sigma_{\rm sm})
-\frac{s^2}{2}\operatorname{Var}_{\rm patch}\Sigma\\
&\qquad+\frac{is^3(\Delta z_{\rm eff})^2}{3}\int dz\,\rho_{\rm sm}^3+\cdots.
\end{split}
\label{eq:patch-gaussian-expansion}
\end{equation}
For fixed $s\sqrt{\operatorname{Var}_{\rm patch}\Sigma}$, a sufficient condition for the expansion to be uniform along the line of sight is
\begin{equation}
\frac{\Delta z_{\rm eff}\max_z\rho_{\rm sm}(\bm\xi,z)}
{\sqrt{\operatorname{Var}_{\rm patch}\Sigma(\bm\xi)}}\ll1.
\label{eq:patch-gaussian-condition}
\end{equation}
Since $\sum_j a_j^3\leq(\max_j a_j)\sum_j a_j^2$, this condition suppresses the cubic term, and higher orders are bounded in the same way. The remaining Gaussian integral gives Eq.~\eqref{eq:patch-gaussian-limit}. Using $\Delta z_{\rm eff}\simeq L_{\rm eff}/N_{\rm eff}$ and Eq.~\eqref{eq:neff}, the contribution from an individual patch satisfies
\begin{equation}
\begin{split}
&\frac{\Delta z_{\rm eff}\rho_{\rm sm}(\bm\xi,z)}
{\sqrt{\operatorname{Var}_{\rm patch}\Sigma(\bm\xi)}}\\
&\qquad\simeq
\frac{L_{\rm eff}(\bm\xi)\rho_{\rm sm}(\bm\xi,z)}
{\Sigma_{\rm sm}(\bm\xi)}
\frac{1}{\sqrt{N_{\rm eff}(\bm\xi)}}.
\end{split}
\label{eq:patch-gaussian-profile}
\end{equation}
The maximum of the profile factor over $z$ determines the threshold in Eq.~\eqref{eq:patch-gaussian-condition}. This Gaussian limit describes fluctuations around the mean on the scale of the standard deviation.

Independence of the exponential contributions gives $\langle(\delta\Sigma)^3\rangle_{\rm patch}=2\sum_j a_j^3$ and $\langle(\delta\Sigma)^4\rangle_{\rm patch}-3(\operatorname{Var}_{\rm patch}\Sigma)^2=6\sum_j a_j^4$. Converting the sums to profile integrals gives the skewness and excess kurtosis,
\begin{equation}
\frac{\langle(\delta\Sigma)^3\rangle_{\rm patch}}
{\left(\operatorname{Var}_{\rm patch}\Sigma\right)^{3/2}}
\simeq
\frac{2}{\sqrt{N_{\rm eff}}}
\frac{\left(\int dz\,\rho_{\rm sm}\right)
\left(\int dz\,\rho_{\rm sm}^3\right)}
{\left(\int dz\,\rho_{\rm sm}^2\right)^2},
\label{eq:patch-skewness}
\end{equation}
\begin{equation}
\frac{\langle(\delta\Sigma)^4\rangle_{\rm patch}}
{\left(\operatorname{Var}_{\rm patch}\Sigma\right)^2}-3
\simeq
\frac{6}{N_{\rm eff}}
\frac{\left(\int dz\,\rho_{\rm sm}\right)^2
\left(\int dz\,\rho_{\rm sm}^4\right)}
{\left(\int dz\,\rho_{\rm sm}^2\right)^3}.
\label{eq:patch-excess-kurtosis}
\end{equation}
All integrals are along the same line of sight.

For the SIS at fixed $\xi>0$, the profile integrals for $p>1/2$ are
\begin{equation}
\int dz\,\rho_{\rm sm}^p(\xi,z)=
\left(\frac{\sigma_v^2}{2\pi G}\right)^p
\xi^{1-2p}\sqrt{\pi}\,
\frac{\Gamma(p-\tfrac12)}{\Gamma(p)}.
\label{eq:sis-column-integrals}
\end{equation}
The cases $p=1,2$ yield Eq.~\eqref{eq:neff-sis}. The profile ratios in Eqs.~\eqref{eq:patch-skewness} and~\eqref{eq:patch-excess-kurtosis} are $3/2$ and $5/2$, respectively, giving Eq.~\eqref{eq:sis-patch-nongaussian}. Also, $L_{\rm eff}\rho_{\rm sm}(\xi,z)/\Sigma_{\rm sm}=2\xi^2/(\xi^2+z^2)$ has maximum $2$, so Eq.~\eqref{eq:patch-gaussian-condition} reduces to $2/\sqrt{N_{\rm eff}}\ll1$.

To evaluate Eq.~\eqref{eq:patch-characteristic} for the SIS, use $\partial_b\int dz\,\ln[1+b/(\xi^2+z^2)]=\pi/\sqrt{\xi^2+b}$ and integrate from $b=0$ to $b=is\Delta z_{\rm eff}\sigma_v^2/(2\pi G)$. This gives
\begin{equation}
\ln\left\langle e^{-is\Sigma}\right\rangle_{\rm patch}^{\rm SIS}
\simeq N_{\rm eff}
\left[1-\sqrt{1+\frac{2is\Sigma_{\rm sm}}{N_{\rm eff}}}\right],
\label{eq:sis-patch-characteristic}
\end{equation}
with the square root continued from $+1$ at $s=0$. Fourier inversion yields
\begin{equation}
\begin{split}
p_{\rm patch}^{\rm SIS}(\Sigma|\xi)\simeq{}&
\sqrt{\frac{N_{\rm eff}\Sigma_{\rm sm}}{2\pi\Sigma^3}}\\
&\times\exp\left[
-\frac{N_{\rm eff}(\Sigma-\Sigma_{\rm sm})^2}
{2\Sigma\Sigma_{\rm sm}}\right],
\qquad \Sigma>0.
\end{split}
\label{eq:sis-patch-pdf}
\end{equation}
The transform can be verified using $\int_0^\infty dx\,x^{-3/2}e^{-a x-b/x}=\sqrt{\pi/b}\,e^{-2\sqrt{ab}}$ for $a,b>0$, followed by analytic continuation.

For comparison with the continuous correlations of $\Psi$, Wick contraction gives two cyclic products of amplitude covariances in the connected three-point density correlation. In the slowly varying limit, each contains the relative coordinate integral $\int du\,dv\,e^{-[u^2+v^2+(u-v)^2]/(2\ell_c^2)}=2\pi\ell_c^2/\sqrt{3}$. Consequently,
\begin{equation}
\begin{split}
\langle(\delta\Sigma)^3\rangle_{\Psi}
&\simeq\frac{4\pi\ell_c^2}{\sqrt{3}}
\int dz\,\rho_{\rm sm}^3(z),\\
\langle(\delta\Sigma)^3\rangle_{\rm patch}
&\simeq2\pi\ell_c^2\int dz\,\rho_{\rm sm}^3(z).
\end{split}
\label{eq:patch-third-moment-comparison}
\end{equation}
For the fourth moment, subtracting $3(\operatorname{Var}_{\Psi}\Sigma)^2$ leaves six cyclic products. Each has a relative coordinate integral $\sqrt{2}\pi^{3/2}\ell_c^3$, giving
\begin{equation}
\begin{split}
&\langle(\delta\Sigma)^4\rangle_{\Psi}-3(\operatorname{Var}_{\Psi}\Sigma)^2\\
&\qquad\simeq6\sqrt{2}\pi^{3/2}\ell_c^3\int dz\,\rho_{\rm sm}^4(z).
\end{split}
\label{eq:patch-fourth-moment-comparison}
\end{equation}
The corresponding patch result is $6\pi^{3/2}\ell_c^3\int dz\,\rho_{\rm sm}^4(z)$. Since the leading variances agree, retaining the continuous correlations increases the skewness and excess kurtosis by factors $2/\sqrt{3}$ and $\sqrt{2}$, respectively, at this order, with respect to the coherent patch results of Eq.~\eqref{eq:sis-patch-nongaussian}.

\subsection{Numerical construction of surface density realizations}
\label{app:uldm-kl}

Under the local transverse approximation of Sec.~\ref{sec:uldm-stats:kl}, the density weighted line-of-sight kernel at the reference position $\bm\xi_0$ is
\begin{equation}
K(z,z')=
\sqrt{\rho_{\rm sm}(\bm\xi_0,z)\rho_{\rm sm}(\bm\xi_0,z')}
e^{-(z-z')^2/(2\ell_c^2)}.
\label{eq:kl-kernel}
\end{equation}
The eigenvalue problem $\int dz'\,K(z,z')u_n(z')=\lambda_nu_n(z)$ determines the weights in Eq.~\eqref{eq:kl-sum}.

The transverse factor in Eq.~\eqref{eq:psi-inhomogeneous} fixes the mode correlations,
\begin{equation}
\begin{split}
\langle\eta_n(\bm\xi)\eta_m^*(\bm\xi')\rangle
&=\delta_{nm}e^{-|\bm\xi-\bm\xi'|^2/(2\ell_c^2)},\\
\langle\eta_n(\bm\xi)\eta_m(\bm\xi')\rangle&=0.
\end{split}
\label{eq:eta-covariance}
\end{equation}
The transverse power spectrum is
\begin{equation}
P_\eta(k_\perp)=2\pi\ell_c^2e^{-\ell_c^2k_\perp^2/2},
\qquad
\int\frac{d^2k_\perp}{(2\pi)^2}P_\eta(k_\perp)=1.
\label{eq:eta-power-spectrum}
\end{equation}
We generate the mode fields using independent complex Gaussian Fourier coefficients with this spectrum.

For quadrature nodes $z_i$ and positive weights $w_i$, including any coordinate Jacobian, define $\rho_i=\rho_{\rm sm}(\bm\xi_0,z_i)$. The longitudinal integral operator is represented by the symmetric matrix
\begin{equation}
M_{ij}=
\sqrt{w_i\rho_i}\,
e^{-(z_i-z_j)^2/(2\ell_c^2)}
\sqrt{w_j\rho_j}.
\label{eq:kl-quadrature}
\end{equation}
Its eigenvalues $\widehat\lambda_n$ approximate the continuum weights. The basic implementation retains a finite set of modes and applies a common normalization,
\begin{equation}
\lambda_n^{\rm used}=
\Sigma_{\rm sm}\,
\frac{\widehat\lambda_n}
{\sum_{m\in{\rm retained}}\widehat\lambda_m},
\qquad n\in{\rm retained}.
\label{eq:kl-weight-normalization}
\end{equation}
The weights are fixed before sampling the random fields. With unit variance $\eta_n$, the ensemble mean is $\Sigma_{\rm sm}$ and the variance is $\sum_n(\lambda_n^{\rm used})^2$. Convergence requires this variance to approach Eq.~\eqref{eq:surface-covariance-exact} at coincident transverse positions as the quadrature and retained modes are refined.

\section{Convergence and comparison with \texttt{GLoW}}
\label{app:glow}
\begin{figure*}[t]
\centering
\includegraphics[width=0.98\textwidth]{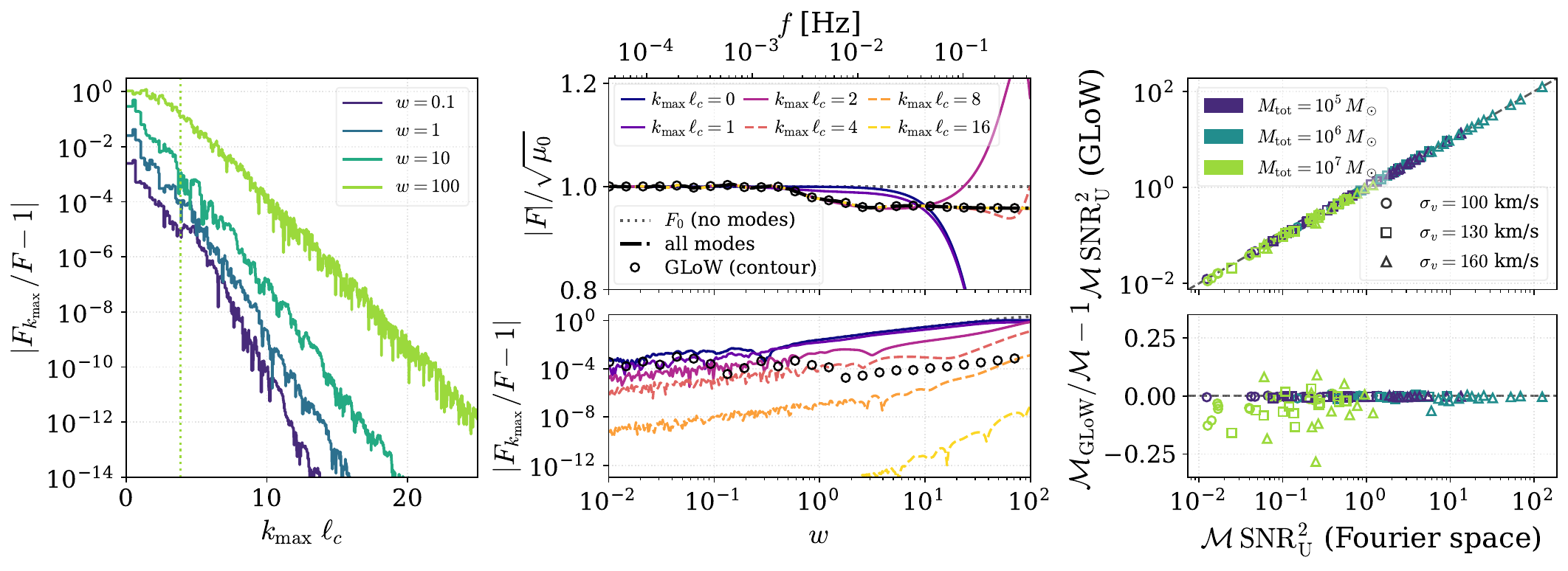}
\caption{Convergence of the Fourier-space evaluation of $F(w)$, Eq.~\eqref{eq:fourier-fresnel}, and its comparison with \texttt{GLoW}. \emph{\textbf{Left and middle.}} One granule realization of the $\sigma_v=160$~km/s host at $m_\phi=10^{-21}$~eV ($\kappa_0=0.363$, Type~I, $\sqrt{\mu_0}=1.91$; $256^{2}$ grid, $L_{\rm box}=13.7\,\ell_c$, $\bm y=(0.3,0)$). \emph{\textbf{Left.}} Relative error of the sum truncated at $|\bm k|\le k_{\max}$, $F_{k_{\max}}$, against the sum over every mode on the grid, at four frequencies; the dotted line marks $w\max|\nabla\psi|$ at $w=100$, the broadening of the mode spectrum by the granule deflection. \emph{\textbf{Middle, top.}} $|F|/\sqrt{\mu_0}$ for the unperturbed lens $F_0$ (dotted) and for increasing $k_{\max}$, compared with the sum over all modes and with \texttt{GLoW}'s contour integration of the same lens (open circles); the top axis gives the GW frequency. \emph{\textbf{Middle, bottom.}} $|F_{k_{\max}}/F-1|$ for the same truncations and for \texttt{GLoW} (open circles), after referring \texttt{GLoW} to the same phase. \emph{\textbf{Right, top.}} $\mathcal{M}\,{\rm SNR}_{\rm U}^{2}$ from \texttt{GLoW} against the Fourier-space evaluation, for every realization on which the contour method completes in the three Type~I ensembles and for the three MBHB sources; the dashed line marks equality. \emph{\textbf{Right, bottom.}} Their fractional difference.}
\label{fig:fourier-convergence}
\end{figure*}

In this appendix we test the Fourier-space evaluation of $F(w)$ described in Sec.~\ref{sec:methods:fourier}. We first examine how the mode sum in Eq.~\eqref{eq:fourier-fresnel} converges, and then compare its result with the public \texttt{GLoW} code~\cite{VillarrubiaRojo2024GLoW}\footnote{\url{https://glow-astro.github.io/GLoW}}, using its multi-contour integration method of the same lens. Both tests are summarized in Fig.~\ref{fig:fourier-convergence}. For the convergence test we use one granule realization of the $\sigma_v=160$~km/s host, a Type~I image with $\kappa_0=0.363$, and keep only the modes with $|\bm k|\le k_{\max}$.
As the left panel of Fig.~\ref{fig:fourier-convergence} shows, the number of modes required grows with frequency. At low frequency the granule phase is small, $g_w\simeq-iw\psi$, and the spectrum $\hat g_w(\bm k)$ follows the granule power spectrum.

A truncated sum therefore fails first at high frequency (middle panels). In the extreme case $k_{\max}=0$, only the box average of the granule phase survives, $F_0\langle e^{-iw\psi}\rangle_{\rm box}$, and $|F|$ drops well below the baseline once $w\psi$ exceeds a radian. Keeping $|\bm k|\le16/\ell_c$ already reproduces $F$ to $10^{-7}$ for all $w\le100$. Our results keep every mode on the grid, so truncation does not affect them.

For the comparison we run \texttt{GLoW}'s contour integration on the same lens map, with the adaptive sampling of $I(\tau)$ tightened to a tolerance of $3\times10^{-5}$. The method applies only to Type~I images (Sec.~\ref{sec:methods:fourier}). The two codes also measure time from different origins: Eq.~\eqref{eq:fourier-fresnel} from the unperturbed image, \texttt{GLoW} from the lowest-delay image of the perturbed lens. We remove the resulting phase factor $e^{-iw\Delta t_{\rm min}}$ before comparing complex values of $F$; it affects neither $|F|$ nor the mismatch (Sec.~\ref{sec:results:setup}).

As the middle panels show, the two methods agree closely. Over $10^{-2}\le w\le10^{2}$, the range covered by \texttt{GLoW}'s Fourier transform, their relative difference stays at or below $\sim10^{-3}$ (median $2\times10^{-4}$), far below the granule modulation. The same holds for the three Type~I ensembles at $\sigma_v=100$, $130$ and $160$~km/s, where the median difference in $|F|$ is $(1$--$4)\times10^{-5}$ of $\sqrt{\mu_0}$. The right panels compare the distinguishability directly. The ensemble medians of $\mathcal{M}\,{\rm SNR}_{\rm U}^{2}$ agree to better than $1\%$ for the $10^5$ and $10^6\,M_\odot$ binaries, and to within $5$--$12\%$ for the $10^{7}\,M_\odot$ binary.

We conclude that Eq.~\eqref{eq:fourier-fresnel} reproduces the contour method wherever the latter applies. It also returns $F(w)$ for every realization, whereas the contour tracer completes on only $13$, $15$ and $13$ of the $30$ realizations at the three velocity dispersions.

\section{Born-order amplitude fluctuations}
\label{app:Fw:born}

We use the Born approximation to interpret the frequency and macroimage position dependence of the ensemble amplitude fluctuations. To first order in the granule potential, $g_w\simeq-iw\psi$, Eq.~\eqref{eq:fourier-fresnel} reduces to the Born-order result of Ref.~\cite{OguriTakahashi2022AmpPhase}, with their magnification eigenvalues $\mu_{j,i}=1/a_i$. At this order, only the real part of $F/F_0-1$ changes the amplitude. For a statistically homogeneous field, the variance of $|F|/\sqrt{|\mu_0|}$ is given by Eq.~(29) of Ref.~\cite{OguriTakahashi2022AmpPhase},
\begin{equation}
{\rm Var}(w)=w^{2}\!\int\!\frac{d^{2}k}{(2\pi)^{2}}\,P_\psi(\bm k)\,\sin^{2}\!\Big(\frac{Q_{\bm k}}{2w}\Big) .
\label{eq:born-var}
\end{equation}
we call its square root ${\rm std}(F)$. Here $P_\psi(\bm k)$ is the ensemble power spectrum of the granule lensing potential.
For a finite periodic box, the integral in this continuum prediction is replaced by a sum over the allowed wavevectors,
\begin{equation}
{\rm Var}_{\rm box}(w)=w^{2}\sum_{\bm k\neq0}\frac{P_\psi(\bm k)}{L^{2}}\,\sin^{2}\!\Big(\frac{Q_{\bm k}}{2w}\Big) .
\label{eq:var-box}
\end{equation}
which omits all power below the fundamental mode $2\pi/L$.

The kernel $\sin^{2}(Q_{\bm k}/2w)$ interpolates between the two optical regimes. For modes longer than the Fresnel scale, $|Q_{\bm k}|\ll w$, it tends to $(Q_{\bm k}/2w)^{2}$, and Eq.~\eqref{eq:born-var} reduces to the variance of the geometric-optics amplitude perturbation, $\tfrac12{\rm tr}(A^{-1}H_\psi)$ with $(H_\psi)_{ab}=\partial_a\partial_b\psi$, whose term $\partial_2^{2}\psi/(1-2\kappa_0)$ is enhanced near the critical curve. For shorter modes it oscillates about $1/2$, so that their contribution is weighted by $w^{2}P_\psi$ alone and is suppressed by the steep decline of $P_\psi$.

\emph{\textbf{ULDM power spectrum.}} For the ULDM field the integrand is fixed by the host and $\ell_c$ alone. The Poisson equation of Sec.~\ref{sec:methods:fw} gives $P_\psi=4\kappa_0^{2}P_\delta/k^{4}$. The transverse correlation $e^{-r^{2}/\ell_c^{2}}$ of Eq.~\eqref{eq:drho-2pt} gives $P_\delta(k)=P_0\,e^{-k^{2}\ell_c^{2}/4}$ with $P_0=\pi\ell_c^{2}\,{\rm Var}(\delta)$. Hence, in units $\ell_c=1$,
\begin{equation}
{\rm Var}(w)=\frac{\kappa_0^{2}\,{\rm Var}(\delta)\,w^{2}}{\pi}\!\int\! d^{2}k\,\frac{e^{-k^{2}/4}}{k^{4}}\,\sin^{2}\!\Big(\frac{Q_{\bm k}}{2w}\Big) .
\label{eq:born-var-uldm}
\end{equation}
The integrand stays finite as $\bm k\to0$, where $\sin^{2}(Q_{\bm k}/2w)\propto k^{4}$ cancels the $k^{-4}$ of the potential.

\begin{figure*}[t]
\centering \includegraphics[width=0.95\textwidth]{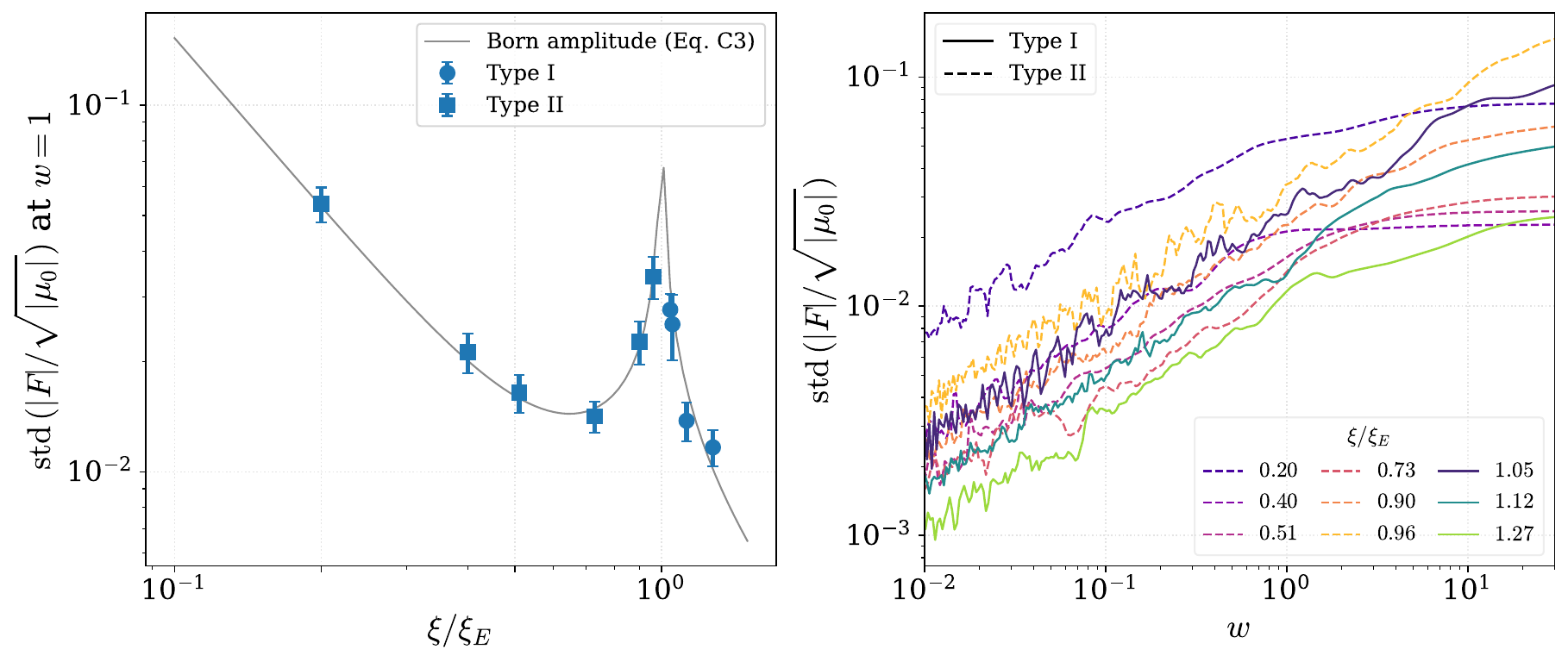}
\caption{Granularity signal ${\rm std}(|F|/\sqrt{|\mu_0|})$ for an SIS host with $\sigma_v=220$~km/s and $m_\phi=10^{-21}$~eV ($\ell_c=8.73$~pc). \emph{\textbf{Left.}}~The signal at $w=1$ versus image position. The points show the standard deviation across independent granule realizations, with $F$ evaluated using Eq.~\eqref{eq:fourier-fresnel}, for Type~I minima (circles) and Type~II saddles (squares). Error bars indicate uncertainty in the estimated standard deviation. The solid curve gives the Born prediction, Eq.~\eqref{eq:born-var-uldm}, using ${\rm rms}(\delta)\propto N_{\rm eff}^{-1/2}$ with the overall normalization determined from the density contrast measured in the ensembles. \emph{\textbf{Right.}}~The signal as a function of $w$, computed from the same numerical ensembles, for Type~I (solid) and Type~II (dashed) images. In both panels, the numerical ensemble standard deviations are divided by $R(w)$ to correct for the bias from the finite periodic box.}
\label{fig:type-i-ii-summary}
\end{figure*}

Equation~\eqref{eq:born-var-uldm} gives the Born prediction shown in the left panel of Fig.~\ref{fig:type-i-ii-summary}, with ${\rm std}(F)_{\rm Born}=\sqrt{\rm Var}$. Since ${\rm rms}(\delta)\propto N_{\rm eff}^{-1/2}$, it gives ${\rm std}(F)\propto\kappa_0N_{\rm eff}^{-1/2}$ times a kernel that depends on $\kappa_0$ only through $A^{-1}$. At fixed $w$ the variance grows as $|\mu_0|$ as $\kappa_0\to1/2$, so that ${\rm std}(F)\propto|\mu_0|^{1/2}$; on the critical curve itself, where $Q_{\bm k}$ grows without bound, the expansion no longer holds. To compare this prediction with the ensemble standard deviations obtained by evaluating the full diffraction integral for the granule realizations, we divide those standard deviations in both panels by $R(w)=[{\rm Var}_{\rm box}(w)/{\rm Var}(w)]^{1/2}$ to correct for the bias from the finite periodic box. For the cases shown, $R$ lies within $4\%$ of unity for $w\gtrsim1$ and falls to $\approx0.3$--$0.5$ at $w=0.1$ for images near the critical curve. We evaluate the ratio using the Born expressions in Eqs.~\eqref{eq:var-box} and~\eqref{eq:born-var}. The agreement in the left panel supports using the Born approximation to interpret the main dependence of the amplitude fluctuations on macroimage position for the cases shown.

%%%%%
% Born sanity check plot
%%%%%

\bibliography{refs}

\end{document}